\documentclass[aps,prd,superscriptaddress,twocolumn,nofootinbib]{revtex4-1}
\usepackage[utf8]{inputenc}
\usepackage{dsfont}
\usepackage{amsfonts}
\usepackage{amsmath}
\allowdisplaybreaks[4]        
\usepackage{amssymb}
\usepackage{euscript}     
\usepackage{braket}
\usepackage{starfont}
\usepackage{color,soul}         
\usepackage{tensor}        
\usepackage{amsthm}
\usepackage{graphicx}
\usepackage{slashed}
\usepackage{leftidx}
\usepackage{subfigure}
\usepackage{bbm}
\definecolor{outerspace}{rgb}{0.25, 0.29, 0.3}
\definecolor{scarlet}{rgb}{1.0, 0.13, 0.0}
\usepackage[header,title,page,titletoc]{appendix}  
\definecolor{princetonorange}{rgb}{1.0, 0.56, 0.0}
\definecolor{WildStrawberry}{rgb}{1.0, 0.26, 0.64}
\definecolor{rossocorsa}{rgb}{0.83, 0.0, 0.0}
\definecolor{navyblue}{rgb}{0.0, 0.0, 0.5}
\usepackage{float}
\usepackage[paper=letterpaper,margin=1in]{geometry}
\usepackage{mathtools}

\DeclareMathAlphabet{\pazocal}{OMS}{zplm}{m}{n}
\newcommand{\req}[1]{(\ref{#1})} 

\newcommand{\bea}{\begin{eqnarray}}

\newcommand{\eea}{\end{eqnarray}}
\newcommand{\ba}{\begin{eqnarray}}
\newcommand{\ea}{\end{eqnarray}}

\newcommand{\be}{\begin{equation}}
\newcommand{\ee}{\end{equation} }
\newcommand{\beqa}{\begin{eqnarray}}
\newcommand{\eeqa}{\end{eqnarray}}
\newcommand{\beqar}{\begin{eqnarray*}}
\newcommand{\eeqar}{\end{eqnarray*}}

\renewcommand{\req}[1]{(\ref{#1})}

\newcommand{\E}{\mathcal{E}}

\makeatletter
\newcommand{\dal}{\mathop{\mathpalette\dal@\relax}}
\newcommand{\dal@}[2]{%
  \begingroup
  \sbox\z@{$\m@th#1\square$}%
  \dimen0=\fontdimen8
    \ifx#1\displaystyle\textfont\else
    \ifx#1\textstyle\textfont\else
    \ifx#1\scriptstyle\scriptfont\else
    \scriptscriptfont\fi\fi\fi3
  \makebox[\wd\z@]{%
    \hbox to \ht\z@{%
      \vrule width \dimen0
      \kern-\dimen0
      \vbox to \ht\z@{
        \hrule height \dimen0 width \ht\z@
        \vss
        \hrule height 2\dimen0
      }%
      \kern-2.5\dimen0
      \vrule width 2.5\dimen0
    }%
  }%
  \endgroup
}
\makeatother

\usepackage{hyperref}
\hypersetup{
    colorlinks,
    citecolor=blue,
    filecolor=navyblue,
    linkcolor=navyblue,
    urlcolor=navyblue
}

\begin{document}

\title{String gravity in $D=4$}
\author{Pablo A. Cano}
\email{pabloantonio.cano@kuleuven.be}
\affiliation{Instituut voor Theoretische Fysica, KU Leuven. Celestijnenlaan 200D, B-3001 Leuven, Belgium. }

\author{Alejandro Ruip\'erez}
\email{alejandro.ruiperez@pd.infn.it}
\affiliation{Dipartimento di Fisica ed Astronomia ``Galileo Galilei", Universit\`a di Padova, Via Marzolo 8, 35131 Padova, Italy}
\affiliation{INFN, Sezione di Padova, 
Via Marzolo 8, 35131 Padova, Italy}


\begin{abstract}

We revisit the four-dimensional theory of gravity that arises from string theory with higher-derivative corrections. By compactifying and truncating the ten-dimensional effective action of heterotic string theory at first order in $\alpha'$, and carefully dealing with field redefinitions, we show that the four-dimensional theory takes the form of an axidilaton model where the scalars couple to the Gauss-Bonnet and Pontryagin densities. Thus, the actual string gravity is a generalization of the well-studied Einstein-dilaton-Gauss-Bonnet and dynamical Chern-Simons models.  Using this action we compute the stringy corrections to the Kerr geometry and we obtain, for the first time, the corrections to the entropy of the Kerr black hole at order $\alpha'^2$. We check that the first law of black hole mechanics is satisfied and discuss several properties of the solution. Our results suggest that there exist black hole solutions with $J>M^2$ and therefore the extremal ratio $J/M^2$ must be modified positively.

\end{abstract}
\maketitle

\section{Introduction}
String theory is thought to provide a quantum theory of gravity and a unified framework for all the forces of Nature. However, it possesses a huge landscape of low-energy effective theories that include all sorts of interactions, many of which we do not have any experimental evidence. One exception to this is gravity itself. Thus, it makes sense to wonder about the stringy prediction for the theory of gravity in our four-dimensional universe.


It has been known since long ago that the gravitational dynamics in string theories is not ruled by the Einstein field equations, but by higher-derivative extensions thereof.
Indeed, the low-energy limit of the different types of string theories can be described by ten-dimensional supergravity actions with higher-derivative corrections \cite{Callan:1985ia,Gross:1986iv,Gross:1986mw,Metsaev:1987zx,Hull:1987pc,Bergshoeff:1988nn,Bergshoeff:1989de}. The compactification of these actions down to four dimensions leads to effective theories with many scalar and vector fields coupled to gravity, but it is interesting to ask what is the minimal truncation one could perform of these theories. Such minimal theory would correspond to the stringy description of pure gravity.
Given that this theory will be different from general relativity (GR), it is simply natural to ask how are the vacuum GR solutions modified by string-theoretical corrections. In particular, one should wonder about the corrections to the arguably most important solution of Einstein's equations: the Kerr black hole.

In this paper we address these questions in the case of heterotic string theory (HST). As is well known, the effective action of HST receives higher-derivative corrections at first order in the string tension, $\alpha'=\ell_{s}^2$ \cite{Callan:1985ia,Gross:1986mw,Metsaev:1987zx,Hull:1987pc,Bergshoeff:1989de}, and famously some of these corrections are  quadratic curvature terms \cite{Zwiebach:1985uq,Deser:1986xr,Duff:1986pq}. 

It is a quite extended lore that, in four dimensions, the effective action of heterotic string theory is captured by the so called Einstein-dilaton-Gauss-Bonnet (EdGB) theory \cite{Boulware:1986dr,Kanti:1995vq,Torii:1996yi,Alexeev:1996vs}. 
\begin{equation}\label{eq:EdGB}
S=\frac{1}{16\pi G}\int d^{4}x\sqrt{|g|}\left[R+\frac{1}{2}(\partial\phi)^2+\alpha e^{-\phi}\mathcal{X}_{4}\right]\, ,
\end{equation}
where $\alpha$ is a parameter with units of length square, and 

\begin{equation}
\mathcal{X}_{4}=R_{\mu\nu \rho\sigma} R^{\mu\nu \rho\sigma}-4R_{\mu\nu}R^{\mu\nu}+R^2\,
\end{equation}
is the Gauss-Bonnet density. This theory has been the subject of very intensive research in the last years (see  refs.~\cite{Moura:2006pz,Guo:2008hf,Maeda:2009uy,Pani:2009wy,Kleihaus:2011tg,Ayzenberg:2014aka,Maselli:2015tta,Kleihaus:2015aje,Kokkotas:2017ymc} for for its black hole solutions), but we would like to revisit the claim about the stringy origin of this model. 


In fact,  the EdGB theory cannot be the complete answer for the low-energy effective action of HST for an important reason: it lacks an axion field. 
In HST, the field strength of the Kalb-Ramond 2-form satisfies the Bianchi identity $dH=\alpha' R \wedge R$, and hence one cannot truncate this field, which must necessarily be a part of the gravitational sector together with the metric and the dilaton.
In fact, this has inspired another family of well-studied models known as dynamical Chern-Simons theory \cite{Campbell:1990fu,Alexander:2009tp}, which can be written as 
\begin{equation}\label{eq:dCS}
S=\frac{1}{16\pi G}\int d^{4}x\sqrt{|g|}\left[R+\frac{1}{2}(\partial\varphi)^2+\beta \varphi \tilde R^2\right]\, ,
\end{equation}
where in this case

\begin{equation}
\tilde R^2=\frac{1}{2}\epsilon^{\mu\nu\alpha\beta}R_{\mu\nu\rho\sigma}\tensor{R}{_{\alpha\beta}^{\rho\sigma}}
\end{equation}
is the Pontryagin density. Black hole solutions in this theory are also known \cite{Yunes:2009hc,Konno:2009kg,Yagi:2012ya,Stein:2014xba,McNees:2015srl,Delsate:2018ome}.

Thus, it is clear that none of the two models above, EdGB and dCS gravities, can be, by themselves, the theory coming from the heterotic string effective action in four dimensions. On the other hand, it is not clear why one should not have other types of higher-derivative terms besides quadratic-curvature ones, such as $(\partial\varphi)^4$, $(\partial\phi)^2(\partial\varphi)^2$, etc. 

Therefore, the first question that we would like to clarify in this paper is what is the precise theory of gravity in four dimensions coming from HST. For that, we will make use of the ten-dimensional effective action at first order in $\alpha'$ as given in \cite{Bergshoeff:1989de}. As we show, the result is an almost (but not exactly) direct generalization of both EdGB and dCS theories. We cover this in section~\ref{sec:action}.

The second question we want to address is that of the stringy corrections to the Kerr metric. It has been known for some time that, at first order in $\alpha'$, the Kerr black hole possesses an axidilatonic hair on account on the non-minimal coupling of these scalars to the curvature \cite{Campbell:1991kz,Campbell:1992hc,Mignemi:1992pm}. However, unlike the case of charged black holes, which receive first-order in $\alpha'$ corrections \cite{Cano:2018qev,Chimento:2018kop,Cano:2018brq,Cano:2019ycn}, the geometry of neutral solutions such as the Kerr black hole is only modified at order $\alpha'^2$. While these effects have been studied in the context of EdGB and dCS gravities (see the references above), the corrections to the Kerr metric in the actual string gravity model have been mostly ignored. As we show, the four-dimensional stringy action at first order in $\alpha'$ can be consistently used to obtain the corrected Kerr metric at order $\alpha'^2$. Thus, in section~\ref{sec:kerrbh} we obtain the corrections to the Kerr geometry expressed analytically as a power series in the spin. We discuss in detail the thermodynamic properties of these rotating black holes, and we compute for the first time the $\alpha'$ corrections to the entropy of the Kerr black hole by using Wald's formula. As an important test of our computations, we check that the first law of black hole mechanics is satisfied. 

Finally, we provide some concluding remarks and discuss possible future directions in section~\ref{sec:conclusions}. \\

\textbf{Note on conventions:}
We follow the conventions of \cite{Ortin:2015hya}: the metric has mostly minus signature and the Riemann tensor is defined by
\begin{equation}
[\nabla_{\mu},\nabla_{\nu}]\xi^{\rho}=\tensor{R}{_{\mu\nu\sigma}^{\rho}}\xi^{\sigma}\, .
\end{equation}
The Ricci tensor is defined in the usual way, $R_{\mu\nu}=\tensor{R}{_{\mu\rho\nu}^{\rho}}$.

\section{Heterotic superstring effective action in four dimensions}\label{sec:action}
The first-order in $\alpha'$ corrections to the heterotic string effective action are fully understood \cite{Gross:1986mw,Metsaev:1987zx,Hull:1987pc,Bergshoeff:1988nn,Bergshoeff:1989de}. Here we will use the action given by ref.~\cite{Bergshoeff:1989de}, which is obtained from the supersymmetrization of the Lorentz-Chern-Simons terms. The equivalence of this result with the ones obtained from string amplitudes was determined in \cite{Chemissany:2007he}. 

Our starting point is the heterotic superstring effective action at first order in $\alpha'$,

\begin{align}\notag
\hat{S}=&\frac{g_s^2}{16\pi G_{N}^{(10)}}\int d^{10}x\sqrt{|\hat g|}e^{-2\hat\phi}\bigg\{\hat R-4(\partial \hat\phi)^2+\frac{1}{12}\hat H^2\\
&+\frac{\alpha'}{8}\hat R_{(-)\mu\nu ab}\hat R_{(-)}^{\mu\nu ab}+\mathcal{O}(\alpha'^3)\bigg\}\, ,
\label{heterotic1}
\end{align}
where we are already truncating away all of the gauge fields. In this action, $\hat R_{(-)}$ is the curvature of the torsionful spin connection 
\begin{equation}\label{Omegadef}
\tensor{\Omega}{_{(-)}^{a}_{b}}=\tensor{\omega}{^{a}_{b}}-\frac{1}{2}\tensor{H}{_{\mu}^{a}_{b}}dx^{\mu}\, ,
\end{equation}
where $a,b$ are Lorentz indices and $\tensor{\omega}{^{a}_{b}}$ is the usual spin connection. The curvature $\hat R_{(-)}$ can be written in terms of the Riemannian curvature $\hat R$ and in terms of $\hat H$ as follows, 
\begin{equation}\label{Rminus}
\tensor{\hat R}{_{(-)\mu\nu}^{\rho}_{\sigma}}=\tensor{\hat R}{_{\mu\nu}^{\rho}_{\sigma}}-\hat{\nabla}_{[\mu}\tensor{\hat H}{_{\nu]}^{\rho}_{\sigma}}-\frac{1}{2}\tensor{\hat H}{_{[\mu|}^{\rho}_{\alpha}}\tensor{\hat H}{_{|\nu]}^{\alpha}_{\sigma}}\, .
\end{equation}
On the other hand, the 3-form field strength is defined as 
\begin{equation}\label{Hdef}
\hat H=d\hat B+\frac{\alpha'}{4}\omega^{L}_{(-)}\, ,
\end{equation}
where $\hat B$ is the Kalb-Ramond 2-form and $\omega^{L}_{(-)}$ is the Lorentz-Chern-Simons 3-form of the torsionful spin connection. Note that the relation \req{Omegadef} implies that $\hat H$ is defined in a recursive way that produces implicitly an infinite tower of $\alpha'$ corrections according to \req{Hdef} \cite{Nilsson:1986md}. Also note that, due to the Chern-Simons term, $\hat H$ satisfies the Bianchi identity,
\begin{equation}\label{Bianchi}
d\hat H=\frac{\alpha'}{4} \tensor{\hat R}{_{(-)}^{a}_{b}}\wedge\tensor{\hat R}{_{(-)}^{b}_{a}}\, .
\end{equation}
Finally, the string coupling constant $g_s$ is related to the asymptotic vacuum expectation value of the dilaton according to $g_s=e^{\langle\hat\phi_{\infty}\rangle}$, while the ten-dimensional Newton's constant reads $G_{N}^{(10)}=8\pi^6g_{s}^2\ell_{s}^8$. 

Our goal is then to find the simplest compactification and truncation of this theory down to four dimensions and to express it in the most compact or appealing way. We observe that the minimal consistent truncation we can perform consists in considering a direct product compactification on a six-torus, $\mathcal{M}_{4}\times \mathbb{T}^{6}$. This is, the metric takes the form,
\begin{equation}
d\hat s^2=d\bar s^2+dz^{i}dz^{i}\, ,\quad i=1,..., 6\, ,
\end{equation}
where $d\bar s^2$ is the four-dimensional metric in the string frame and the coordinates $z_{i}\sim z_{i}+2\pi \ell_{s}$ parametrize the six-torus. Here we are taking all of the Kaluza-Klein vectors and scalars to be trivial. At the same time, the Kalb-Ramond 2-form and its 3-form field strength only have four-dimensional components, a fact that we express by $\hat B=B$, $\hat H=H$.  That this is a consistent truncation follows from the fact that this compactification ansatz solves all of the ten-dimensional equations of motion once the lower-dimensional ones are satisfied, as one can check from direct inspection.\footnote{Alternatively, one can be convinced of this by examining the full toroidal compactification of HST, whose action is provided in \cite{Ortin:2020xdm}. One can check that truncating all of the vectors and scalars, except for the dilaton, is consistent.}

This trivial compactification gives rise to formally the same theory as \req{heterotic1} but in four dimensions and with a Newton's constant $G_{N}^{(4)}=G_{N}^{(10)}/(2\pi \ell_{s})^6$. 
In order to express this theory in a more appropriate form, we first dualize the Kalb-Ramond 2-form into a scalar field. Let us note that, in four dimensions, the Bianchi identity \req{Bianchi} can be expressed as

\begin{equation}
\frac{1}{3!}\epsilon^{\mu\nu\rho\sigma}\bar{\nabla}_{\mu}H_{\nu\rho\sigma}+\frac{\alpha'}{8}\tensor{\bar R}{_{(-)\mu\nu\rho\sigma}} \tensor{\tilde {\bar R}}{_{(-)}^{\mu\nu\rho\sigma}}=0\, ,
\end{equation}
where 

\begin{equation}
\tensor{\tilde {\bar R}}{_{(-)}^{\mu\nu\rho\sigma}}=\frac{1}{2}\epsilon^{\mu\nu\alpha\beta}\tensor{\bar R}{_{(-)}_{\alpha\beta}^{\rho\sigma}}\, .
\end{equation}
Then, we can promote $H$ to be the dynamical field instead of $B$ by introducing the Bianchi identity in the action together with a Lagrange multiplier $\varphi$. After integration by parts, we are left with the action. 

\begin{align}\notag
\bar{S}&=\frac{1}{16\pi G_{N}^{(4)}}\int d^{4}x\sqrt{|\bar g|}\bigg\{e^{-2(\hat\phi-\hat\phi_{\infty})}\Big(\bar R-4(\partial \hat\phi)^2\\&+\frac{1}{12} H^2\Big)-\frac{1}{3!}H_{\mu\nu\rho}\epsilon^{\mu\nu\rho\sigma}\partial_{\sigma}\varphi+\frac{\alpha'}{8}\mathcal{L}_{R^2}+\mathcal{O}(\alpha'^3)\bigg\}\, ,
\label{heterotic2}
\end{align}
where 
\begin{equation}\label{LR2}
\mathcal{L}_{R^2}=e^{-2(\hat\phi-\hat\phi_{\infty})}\bar R_{(-)\mu\nu\rho\sigma}\bar R_{(-)}^{\mu\nu\rho\sigma}-\varphi \tensor{\bar R}{_{(-)\mu\nu\rho\sigma}} \tensor{\tilde {\bar R}}{_{(-)}^{\mu\nu\rho\sigma}}\, .
\end{equation}
Now the variation of this action with respect to $\varphi$ yields the Bianchi identity of $H$, while variation with respect to $H$ yields a relation that allows one to remove $H$ in terms of $\varphi$. However, in this case the dualization process is not so straightforward as the Lagrangian has a non-linear dependence on $H$ through $\mathcal{L}_{R^2}$. In fact, we get the following equation from the variation of $H$: 
\begin{equation}
e^{-2(\hat\phi-\hat\phi_{\infty})}\frac{1}{6}H_{\mu\nu\rho}-\frac{1}{6}\epsilon_{\mu\nu\rho\sigma}\bar\nabla^{\sigma}\varphi+\frac{\alpha'}{8}\frac{\delta \mathcal{L}_{R^2}}{\delta H^{\mu\nu\rho}}=0\, .
\end{equation}
In order to solve it, we expand $H$ in a series in $\alpha'$, 

\begin{equation}\label{Hexpansion}
H=H^{(0)}+\alpha' H^{(1)}+\alpha'^2 H^{(2)}+\mathcal{O}(\alpha'^3)\, ,
\end{equation}
and we get the following result for the first two terms,

\begin{align}\label{H0}
H^{(0)}_{\mu\nu\rho}&=e^{2(\hat\phi-\hat\phi_{\infty})}\epsilon_{\mu\nu\rho\sigma}\bar\nabla^{\sigma}\varphi\, ,\\
H^{(1)}_{\mu\nu\rho}&=-e^{2(\hat\phi-\hat\phi_{\infty})}\frac{3}{4}\frac{\delta \mathcal{L}_{R^2}}{\delta H^{\mu\nu\rho}}\Big|_{H^{(0)}}\, .
\end{align}

We then have to plug $H=H(\varphi)$ back into the action so that we eliminate the 3-form in terms of $\varphi$. Remarkably, when \req{Hexpansion} is inserted in \req{heterotic2}, we observe that no additional $\mathcal{O}(\alpha')$ terms are generated, and the action reads simply 

\begin{align}\notag
\bar{S}=&\frac{1}{16\pi G_{N}^{(4)}}\int d^{4}x\sqrt{|\bar g|}\bigg\{e^{-2(\hat\phi-\hat\phi_{\infty})}\Big(\bar R-4(\partial \hat\phi)^2\Big)\\
&+\frac{1}{2}e^{2(\hat\phi-\hat\phi_{\infty})}(\partial\varphi)^2+\frac{\alpha'}{8}\mathcal{L}_{R^2}\Big|_{H^{(0)}}+\mathcal{O}(\alpha'^2)\bigg\}\, .
\label{heterotic3}
\end{align}
We note, however, that the dualization introduces ${\cal O}(\alpha'^2)$ terms which were not present in the original action \eqref{heterotic1}. These terms are given in Appendix~\ref{app:app:alpha2}, where it is also argued that they become of order ${\cal O}(\alpha'^4)$ when the scalars are of order ${\cal O}(\alpha')$. This is relevant for computing corrections to vacuum solutions of GR, as we shall discuss later.
Then, we have to evaluate the four-derivative term $\mathcal{L}_{R^2}$ by substituting the value of $H^{(0)}$ in \req{H0}, for which we also have to use \req{Rminus}. After a somewhat lengthy computation in which we make use of the properties of the Levi-Civita symbol, we find the following answer,

\begin{equation}
\begin{aligned}
\mathcal{L}_{R^2}\Big|_{H^{(0)}}&=e^{-2(\hat\phi-\hat\phi_{\infty})}\Big(\bar R_{\mu\nu \rho\sigma}\bar R^{\mu\nu \rho\sigma}+6\bar G_{\mu\nu}A^{\mu}A^{\nu}\\
&+\frac{7}{4}A^4-2\bar\nabla_{\mu}A_{\nu}\bar\nabla^{\mu}A^{\nu}-(\bar\nabla_{\mu}A^{\mu})^2\Big)\\
&-\varphi \tensor{\bar R}{_{\mu\nu\rho\sigma}} \tensor{\tilde {\bar R}}{^{\mu\nu \rho\sigma}}+\text{total derivatives}\, ,
\end{aligned}
\end{equation}
where we have introduced $A_{\mu}=e^{2(\hat\phi-\hat\phi_{\infty})}\partial_{\mu}\varphi$. Here $\bar R_{\mu\nu \rho\sigma}$ is the standard Riemann tensor of the string frame metric, while $\bar G_{\mu\nu}$ is the Einstein tensor. Now we have to transform our theory to the (modified) Einstein frame. This is achieved by rescaling the metric as 
\begin{equation}\label{conformalrescaling}
\bar g_{\mu\nu}=e^{2(\hat\phi-\hat\phi_{\infty})} g_{\mu\nu}\, .
\end{equation}
The effect of this conformal rescaling on the two-derivative Lagrangian is well known:
\begin{equation}
\begin{aligned}
\sqrt{|\bar g|}\mathcal{L}_{2}&=\sqrt{|g|}\Big(R+2(\partial\hat\phi)^2+\frac{1}{2}e^{4(\hat\phi-\hat \phi_{\infty})}(\partial \varphi)^2\Big) \, .
\end{aligned}
\end{equation}

On the other hand, it takes some computations to obtain the transformation of the four-derivative Lagrangian under this conformal rescaling. After making use of the transformation rules of the Riemann tensor and of the covariant derivative, and integrating by parts multiple times, we can express the result as follows, up to total derivatives that we omit:

\begin{equation}
\begin{aligned}
&\sqrt{|\bar g|}\mathcal{L}_{R^2}\Big|_{H^{(0)}}=\sqrt{|g|}\bigg[e^{-2(\hat\phi-\hat\phi_{\infty})}\Big(\tensor{R}{_{\mu\nu\rho\sigma}} \tensor{R}{^{\mu\nu \rho\sigma}}\\
&+4R^{\mu\nu}(4\partial_{\mu}\hat\phi\partial_{\nu}\hat\phi+A_{\mu}A_{\nu})\\
&+R(4\nabla^2\hat\phi-4(\partial\hat\phi)^2-3 A^2)+12(\partial\hat\phi)^4+12(\nabla^2\hat\phi)^2\\
&+\frac{7}{4}A^4-12(\partial_{\mu}\hat\phi A^{\mu})^2-2A^2(\partial\hat\phi)^2-8A^2\nabla^2\hat\phi\\
&-16\partial_{\mu}\hat\phi A^{\mu}\nabla_{\alpha}A^{\alpha}-3(\nabla_{\alpha}A^{\alpha})^2\Big)-\varphi \tensor{R}{_{\mu\nu\rho\sigma}} \tensor{\tilde {R}}{^{\mu\nu \rho\sigma}}\bigg] \, .
\end{aligned}
\end{equation}
This result is not very illuminating, but we can massage it a bit further. Let us first rewrite this expression in terms of the Gauss-Bonnet density, which is defined by
\begin{equation}
\mathcal{X}_{4}=R_{\mu\nu \rho\sigma} R^{\mu\nu \rho\sigma}-4R_{\mu\nu}R^{\mu\nu}+R^2\, .
\end{equation}
Thus, we can simply replace the Riemann squared term by the Gauss-Bonnet density using $R_{\mu\nu \rho\sigma} R^{\mu\nu \rho\sigma}=\mathcal{X}_{4}+4R_{\mu\nu}R^{\mu\nu}-R^2$. Then, we can express the result as

\begin{equation}
\begin{aligned}
\sqrt{|\bar g|}\mathcal{L}_{R^2}\Big|_{H^{(0)}}&=\sqrt{|g|}\Big(e^{-2(\hat\phi-\hat\phi_{\infty})}\mathcal{X}_{4}\\
&-\varphi \tensor{R}{_{\mu\nu\rho\sigma}} \tensor{\tilde {R}}{^{\mu\nu \rho\sigma}}+\mathcal{L}'\Big) ,
\end{aligned}
\end{equation}
where we are simply collecting the rest of the terms in $\mathcal{L}'$, whose form, as we have seen, is quite complicated. However, this Lagrangian becomes much more illuminating if we write it in terms of the zeroth-order equations of motion, 
\begin{align}
\mathcal{E}_{\mu\nu}&=R_{\mu\nu}+2\partial_{\mu}\hat\phi\partial_{\nu}\hat\phi+\frac{1}{2}A_{\mu}A_{\nu}\, ,\\
\mathcal{E}_{\hat\phi}&=\nabla^2\hat\phi-\frac{1}{2}A^2\, ,\\
\mathcal{E}_{\varphi}&=\nabla_{\mu}A^{\mu}+2\partial_{\mu}\hat\phi A^{\mu}\, .
\end{align}
After some algebra, we obtain the following result

\begin{equation}\label{Lprime}
\begin{aligned}
\mathcal{L}'&=e^{-2(\hat\phi-\hat\phi_{\infty})}\Big[4\mathcal{E}_{\mu\nu}\mathcal{E}^{\mu\nu}-\mathcal{E}^2+12\mathcal{E}_{\hat\phi}^2+4\E\E_{\hat\phi}\\
&-3\E_{\varphi}^2+2\E_{\hat\phi}(A^2-4(\partial\hat\phi)^2)-4\E_{\varphi}\partial_{\mu}\hat\phi A^{\mu}\Big]\, .
\end{aligned}
\end{equation}
Astoundingly, all of the terms in $\mathcal{L}'$ are proportional to the zeroth-order equations of motion. This means that these terms can be removed via a redefinition of the fields of the form 
\begin{equation}
g_{\mu\nu}\rightarrow g_{\mu\nu} +\alpha' \Delta_{\mu\nu}\, ,\,\,\, \hat\phi\rightarrow \hat\phi+\alpha' \Delta {\hat\phi}\, ,\,\,\, \varphi\rightarrow \varphi+\alpha' \Delta {\varphi}\ .
\end{equation}
These redefinitions introduce $\mathcal{O}(\alpha')$ terms proportional to the zeroth equations of motion, and hence we can use them to cancel all of the terms in $\mathcal{L}'$. We show the explicit redefinitions in the Appendix~\ref{app:app:alpha2}.  Of course, these redefinitions also modify the action at higher orders in $\alpha'$, and in particular they introduce $\mathcal{O}(\alpha'^2)$ corrections. Note that, for the terms in \req{Lprime} that are quadratic in the zeroth order EOM, these  $\mathcal{O}(\alpha'^2)$ corrections generated by the redefinitions are still proportional to the EOM and they can be further removed by an additional redefinition. These will then only contribute at order $\mathcal{O}(\alpha'^3)$. However, the two last terms in \req{Lprime} are only linear in the equations of the scalar fields, and therefore we will introduce non-trivial $\mathcal{O}(\alpha'^2)$ corrections upon removing these terms at first order in $\alpha'$. We discuss these terms in more detail in Appendix~\ref{app:app:alpha2}.\footnote{In any case, we note that, as it happens with the ${\cal O}(\alpha'^2)$ terms generated by the dualization, these terms are proportional to the square of derivatives of the scalars. }

In sum, field redefinitions can be used to set $\mathcal{L}'=0$. Finally, introducing the four-dimensional dilaton as $\phi=2(\hat\phi-\hat\phi_{\infty})$ we have a very elegant form for the heterotic string effective action at first order in $\alpha'$ in four dimensions:\footnote{Let us note that essentially the same action was written in \cite{Campbell:1992hc}. However, that paper starts with an action that seems to lack many of the higher-derivative terms of HST (compare eq.~(1) of that paper with eq.~(3.1) of \cite{Metsaev:1987zx}), so the result of \cite{Campbell:1992hc} does not seem to be justified.} \footnote{In the more usual conventions of mostly ``$+$" signature and Riemann tensor defined by $[\nabla_{\mu},\nabla_{\nu}]\xi^{\rho}=\tensor{R}{^{\rho}_{\sigma\mu\nu}}\xi^{\sigma}$, the only difference in the action would be a change of sign in the kinetic terms of the scalar fields, \textit{e.g.}, $(\partial\phi)^2\rightarrow -(\partial\phi)^2$.}

\begin{align}\notag
S=&\frac{1}{16\pi G_{N}^{(4)}}\int d^{4}x\sqrt{|g|}\bigg\{ R+\frac{1}{2}(\partial\phi)^2+\frac{1}{2}e^{2\phi}(\partial \varphi)^2\\\label{eq:stringyfinal}
&+\frac{\alpha'}{8}\left(e^{-\phi}\mathcal{X}_{4}-\varphi R_{\mu\nu\rho\sigma}\tilde R^{\mu\nu\rho\sigma}\right)+\mathcal{O}(\alpha'^2)\bigg\}\, .
\end{align}
This result deserves some comments. First, observe that it is very non-trivial that the only higher-derivative corrections are given by the Gauss-Bonnet and Pontryagin densities. There are other four higher-derivative operators that could be present in the action and that could not be removed by field redefinitions, namely $(\partial\phi)^4$, $e^{3\phi}(\partial\varphi)^4$, $e^{\phi}(\partial\varphi)^2(\partial\phi)^2$ and $e^{\phi}(\partial_{\mu}\phi\partial^{\mu}\varphi)^2$. We remark that our result is completely general and we are not making any approximation, \textit{e.g.}, we are not neglecting those terms by assuming that the scalar fields are of order $\alpha'$. We simply find that these terms are not present.
	
Second, precisely because we do not have those terms, this action is almost exactly equivalent to an axidilaton model where the curvature 2-form plays the r\^ole of the field strength of the gauge field. In fact, introducing $\tau=\varphi+i e^{-\phi}$, and taking into account that the Gauss-Bonnet density can be written as $\mathcal{X}_{4}=-R_{\mu\nu\rho\sigma}\tilde{\tilde R}^{\mu\nu\rho\sigma}$, where $\tilde{\tilde R}^{\mu\nu\rho\sigma}$ is the double dual of the Riemann tensor, we can write the action in a suggestive way as

\begin{align}\notag
S=&\frac{1}{16\pi G_{N}^{(4)}}\int d^{4}x\sqrt{|g|}\bigg\{ R+\frac{\partial_{\mu}\tau\partial^{\mu}\bar\tau}{2|\text{Im}(\tau)|^2}+\\\label{eq:stringyaxidilaton}
&-\frac{\alpha'}{8}\left(\text{Im}(\tau)R_{\mu\nu\rho\sigma}\tilde{\tilde R}^{\mu\nu\rho\sigma}+\text{Re}(\tau) R_{\mu\nu\rho\sigma}\tilde R^{\mu\nu\rho\sigma}\right)\bigg\}\, .
\end{align}
Thus, except for the presence of $R\tilde{\tilde R}$ instead of Riemann$^2$ (which would be the equivalent of $F^2$), this is essentially like an axidilaton model.\footnote{It is very tempting to wonder if this action possesses an $\mathrm{SL}(2,\mathbb{R})$ symmetry that involves the dualization of the curvature. We will not dwell in this in this paper but we think it would be worth to explore this idea elsewhere.}

Third, note that the scalar fields cannot be truncated, and hence they form part of the gravitational sector. In particular, the axion must be non-trivial whenever $R_{\mu\nu\rho\sigma}\tilde R^{\mu\nu\rho\sigma}\neq 0$. For the special case of spherically-symmetric solutions, the Pontryagin density vanishes, and hence one recovers the solutions of EdGB gravity. However, whenever the Pontryagin density is non-trivial, an axion hair is generated, and the solutions will be different from those of EdGB and dCS theories. This happens of course for rotating black holes, but also, \textit{e.g.}, for linear perturbations around spherical black holes.

Finally, observe that, for corrections over Ricci flat solutions, the scalar fields acquire non-trivial profiles of order $\alpha'$ and these backreact into the geometry at order $\alpha'^2$. The reason for this is that the two quadratic densities are topological, and thus they only contribute to the equations of motion as long as the scalars are non-trivial. Now, since the $\mathcal{O}(\alpha'^2)$ terms in the action \req{eq:stringyfinal} are proportional to derivatives of the scalars (see Appendix~\ref{app:app:alpha2}), these terms actually contribute at higher orders when the scalars have profiles of order $\alpha'$. Therefore, the $\mathcal{O}(\alpha')$ action already captures all of the $\mathcal{O}(\alpha'^2)$ corrections to Ricci-flat solutions, such as the Kerr black hole, that we study in the next section.

\section{The Kerr black hole at order $\alpha'^2$}\label{sec:kerrbh}
After having determined the form of the heterotic string effective action in four dimensions as given in \req{eq:stringyfinal}, our goal now is to compute the $\alpha'$ corrections to the Kerr metric in that theory.
Throughout this section we will work in units of $G_{N}^{(4)}=1$, so one has to bear in mind that every quantity is expressed in Planck units. Observe that the value of Newton's constant in our setup is given by $G_{N}^{(4)}=g_{s}^2\alpha'/8$, and therefore, in Planck units we have $\alpha'=8g_{s}^{-2}$. Hence, if we consider a weakly coupled string theory, $g_{s}<<1$, we have $\alpha'>>1$, and thus the corrections appear much below the Planck scale. 

The first aspect that one notices about the solutions of the theory \req{eq:stringyfinal} is that they typically possess an axidilatonic hair generated by the coupling of the scalars to the quadratic invariants. The existence of this hair in the context of string theory was noted long ago \cite{Campbell:1991kz,Campbell:1992hc,Mignemi:1992pm}, but let us quickly review it. The equations for the dilaton and the axion read
\begin{align}
\nabla^2\phi&=e^{2\phi}(\partial\varphi)^2-\frac{\alpha'}{8}e^{-\phi}\mathcal{X}_{4}+\mathcal{O}(\alpha'^2)\, ,\\
\nabla_{\mu}(e^{2\phi}\nabla^{\mu}\varphi)&=-\frac{\alpha'}{8}R_{\mu\nu\rho\sigma}\tilde R^{\mu\nu\rho\sigma}+\mathcal{O}(\alpha'^2)\, ,
\end{align}
and it is clear these scalars cannot be constant as long as the quadratic curvature invariants do not vanish. We can simplify these equations if we are interested in corrections to solutions that have trivial scalars at zeroth order in $\alpha'$ --- this is, corrections to GR solutions. Note that, since $\phi=2(\hat\phi-\hat\phi_{\infty})$, the vacuum value of the four-dimensional dilaton must be zero according to string theory, while the axion can have an arbitrary vacuum value $\varphi_{\infty}$. Then, since the fluctuations of the scalars with respect to their vacuum values will be of order $\alpha'$, one can reduce their equations to the following,

\begin{eqnarray}
\nabla^2\phi&=&-\frac{\alpha'}{8}\mathcal{X}_{4}+\mathcal{O}(\alpha'^2)\, ,\\
\nabla^2\varphi&=&-\frac{\alpha'}{8}R_{\mu\nu\rho\sigma}\tilde R^{\mu\nu\rho\sigma}+\mathcal{O}(\alpha'^2)\, .
\end{eqnarray}
In the background of the Kerr black hole,  one finds a unique stationary and axisymmetric solution to these equations by fixing the asymptotic values of the scalars to their vacuum values and demanding that they are regular at the horizon.
This solution does not seem to admit a fully analytic expression but one can easily expand it in a series in the spin $\chi$ \cite{Mignemi:1992pm,Pani:2009wy,Yunes:2009hc,Konno:2009kg,Yagi:2012ya,Cano:2019ore}. In particular, let us remark that the axionic hair is vanishing for spherically symmetric solutions, but not for rotating ones. 

While the profile of these scalars can only be obtained analytically through a power series in $\chi$, it is possible to find the exact (in spin) value for the dilaton charge, $Q_{\phi}$, identified by the asymptotic expansion of the dilaton, $\phi= -Q_{\phi}/\rho$ when $\rho\rightarrow\infty$. It turns out that this reads \cite{Prabhu:2018aun,Cano:2019ore}
\begin{equation}
Q_{\phi}=2\pi \alpha' \, T\, ,
\end{equation}
where  $T$ is the Hawking temperature of the black hole. This result is known to hold in general for EsGB gravity with a linear coupling \cite{Prabhu:2018aun}, but it holds for heterotic string theory only at first order in $\alpha'$. Likewise, one can find an exact expression for the dipole moment of the axion --- see \cite{Cano:2019ore}. 

While the axion and the dilaton get corrections at first order in $\alpha'$, one can see that these in turn ``backreact" in the geometry at order $\mathcal{O}(\alpha'^2)$. As we have shown, one can use the action \req{eq:stringyfinal} to consistently compute these corrections in heterotic string theory.  Although the corrections to the Kerr metric associated to either the Gauss-Bonnet \cite{Moura:2006pz,Maeda:2009uy,Pani:2009wy,Kleihaus:2011tg,Ayzenberg:2014aka,Maselli:2015tta,Kleihaus:2015aje,Kokkotas:2017ymc} or the Chern-Simons \cite{Yunes:2009hc,Konno:2009kg,Yagi:2012ya,Stein:2014xba,McNees:2015srl,Delsate:2018ome} sectors have been separately studied in the literature, their joint action has been essentially missed. 

In order to compute the corrections to the Kerr geometry, we will follow our previous work \cite{Cano:2019ore} that studies a more general theory motivated by EFT arguments and offers a systematic method to find the corrected Kerr metric in higher-derivative theories. 
As shown in that reference, one can capture the corrections to the Kerr metric by using the following ansatz,

\begin{widetext}
\begin{align}\notag\label{rotatingmetric}
ds^2=&\left(1-\frac{2 M \rho}{\Sigma}-H_1\right)dt^2+\left(1+H_2\right)\frac{4 M a \rho \sin^2\theta}{\Sigma}dtd\phi\\
&-\left(1+H_3\right)\Sigma\left(\frac{d\rho^2}{\Delta}+d\theta^2\right)-\left(1+H_4\right)\left(\rho^2+a^2+\frac{2 M  \rho a^2\sin^2\theta}{\Sigma}\right)\sin^2\theta d\phi^2\, .
\end{align}
\end{widetext}
where $\Sigma=\rho^2+a^2 \cos^2\theta$ and $\Delta=\rho^2-2M\rho+a^2$, and the functions $H_{i}$ contain the corrections. These functions can be obtained analytically by performing a series expansion in the spin $\chi=a/M$ as
\begin{equation}
H_{i}=\sum_{n=0}^{\infty}\chi^n H_{i}^{(n)}(\rho,x)\, ,
\end{equation}
where each term $H_{i}^{(n)}(\rho,x)$ is simply a polynomial in $1/\rho$ and $x=\cos\theta$. When solving the equations of motion we get several integration constants, but we fix these by imposing asymptotic flatness and that the parameters $M$ and $J=\chi M^2$ still represent the total mass and angular momentum. We show the few first terms of these functions in Appendix~\ref{app:thesolution}, but one can solve the equations algorithmically and for the present work we managed to obtain the solution to order $\chi^{40}$. An analysis of convergence indicates that these series are convergent for $\chi<1$, and to order $\chi^{40}$ they are accurate everywhere outside the horizon for $\chi<0.9$. 
It is to be expected that these series do not converge for $\chi\sim 1$, corresponding to near extremality. As a matter of fact, the metric above is not appropriate to study the corrections to extremal black holes, as the ansatz assumes that extremality happens for $\chi=1$ (when $\Delta$ develops a double root), but this is no longer true in the presence of corrections \cite{Kleihaus:2011tg,Kleihaus:2015aje,Reall:2019sah}. One should use a different approach to study the corrections to extremal or near-extremal black holes.  Nevertheless, we will try to say a few things about the extremal limit by extrapolating our results to higher values of $\chi$.

One could study many properties of these corrected black hole geometries, but here we focus on the thermodynamic properties, which were not completely characterized in \cite{Cano:2019ore} --- in particular, the entropy was not computed. 

Let us start by characterizing the event horizon of \req{rotatingmetric}.
The main advantage of the coordinates used in the metric above is that the location of the horizon is determined by the largest root of $\Delta$,
\begin{equation}
\rho_{+}=M\left(1+\sqrt{1-\chi^2}\right)\, ,
\end{equation}
and therefore its position in terms of $\rho$ is not corrected. The Killing vector that generates the horizon must be of the form 

\begin{equation}
\xi=\partial_{t}+\Omega \partial_{\phi}\, ,
\end{equation}
for certain constant $\Omega$. Demanding that $\xi$ becomes null at $\rho=\rho_{+}$ one finds that $\Omega$ is indeed the angular velocity of the horizon,

\begin{equation}
\Omega=\frac{g_{t\phi}}{|g_{\phi\phi}|}\bigg|_{\rho=\rho_{+}}\, .
\end{equation}

\begin{figure}[t!]
	\begin{center}
		\includegraphics[width=0.49\textwidth]{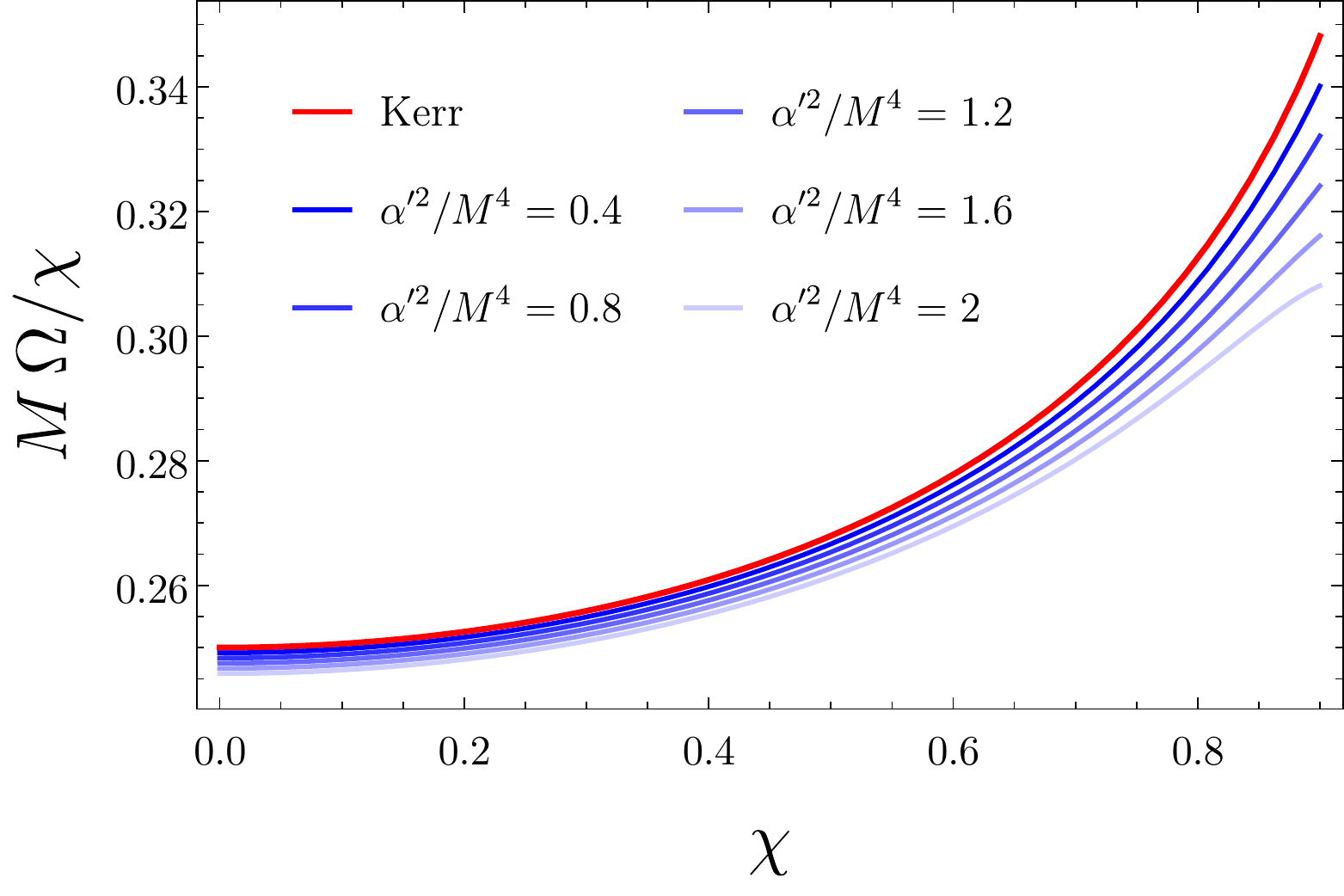}  
		\caption{Angular velocity of rotating black holes at order $\alpha'^2$. We show the dimensionless product $M \Omega$ divided by $\chi$ to better observe the non-linear spin-dependence. The red line corresponds to the Kerr prediction while blue lines of different opacities represent the deviation for several values of the mass relative to the string scale.}
		\label{fig:omega}
	\end{center}
\end{figure}

\begin{figure}[t!]
	\begin{center}
		\includegraphics[width=0.49\textwidth]{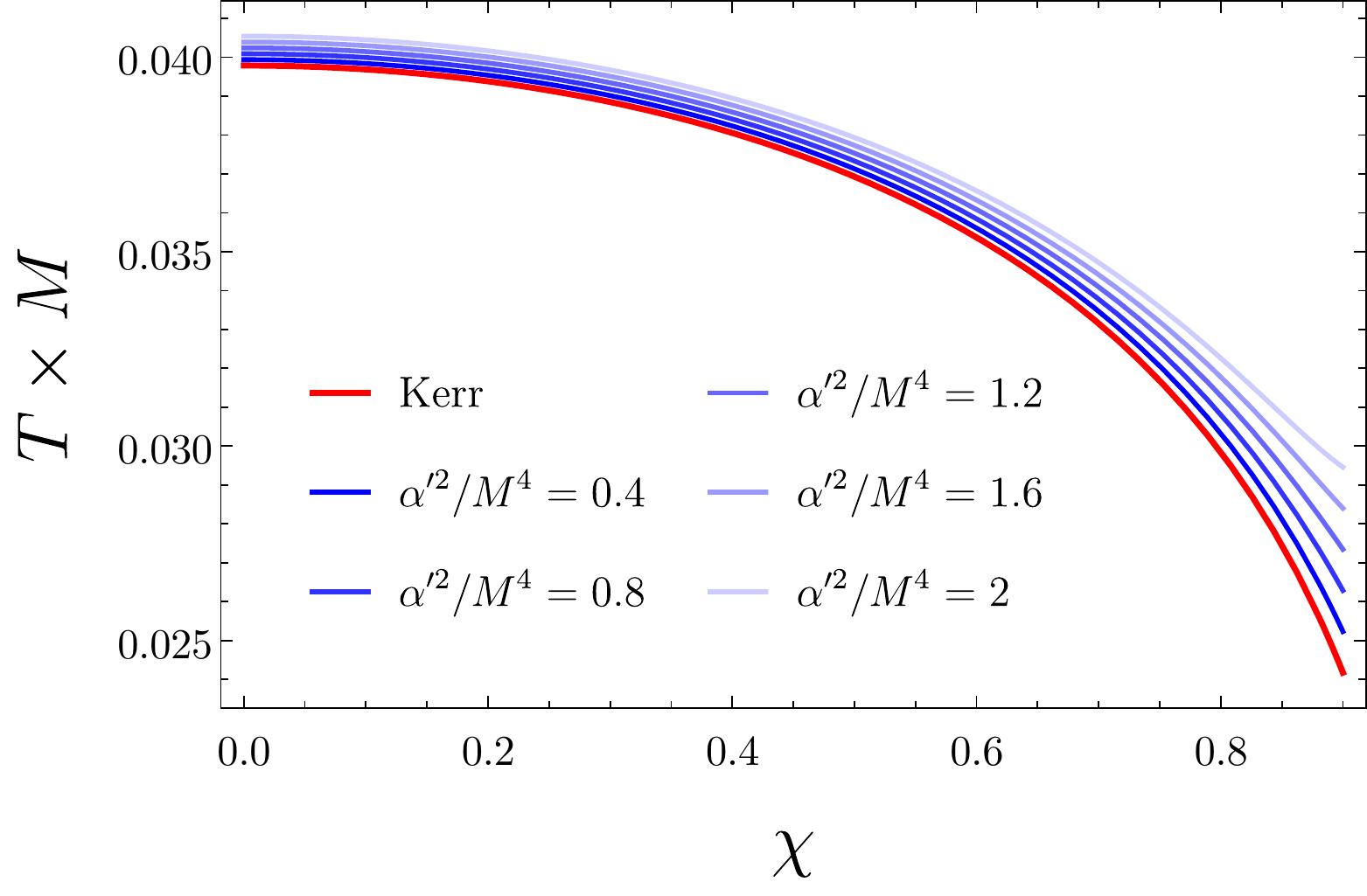}  
		\caption{Temperature of the corrected Kerr black holes as a function of the spin. We show the product $M T$, which is only a function of the spin $\chi$ in Einstein gravity. The red line corresponds to the Kerr prediction while blue lines of different opacities represent the corrections for several values of the mass relative to the string scale.}
		\label{fig:temperature}
	\end{center}
\end{figure}
\noindent
This quantity does receive $\alpha'$ corrections, and the few first terms in the $\chi$ expansion read

\begin{equation}\label{eq:omega}
\begin{aligned}
\Omega&=\frac{\chi}{2M\left(1+\sqrt{1-\chi^2}\right)}-\frac{\chi\alpha'^2}{64M^5}\bigg(\frac{1193 }{8960}\\
&+\frac{173137 \chi^2}{806400}+\frac{9251171 \chi ^4}{35481600}+\mathcal{O}(\chi^6)\bigg)+\mathcal{O}(\alpha'^3)\, .
\end{aligned}
\end{equation}
Notice that the fact that this quantity is actually constant means that the black hole horizon \emph{is} a Killing horizon. This is a quite non-trivial check of the correctness of our solution. Also observe that the corrections to $\Omega$ are negative, so the stringy black holes spin more slowly than Kerr ones. In fig.~\ref{fig:omega} we show $\Omega$ as a function of the spin up to $\chi=0.9$ by using an expansion up to order $\chi^{40}$. The effect of the corrections becomes more relevant as we decrease the ratio $M/\sqrt{\alpha'}$, but they also increase for larger values of the spin.

Once we have determined the Killing vector $\xi$, one can compute its associated surface gravity at the horizon $\kappa$, and consequently obtain the Hawking's temperature of the black hole, which is given by
\begin{equation}
T=\frac{\kappa}{2\pi}\, .
\end{equation}
We remind that the surface gravity is defined by the equation $\xi^{\nu}\nabla_{\nu}\xi^{\mu}=\kappa \xi^{\mu}$, which holds on the horizon. In practice, the computation of $\kappa$ is a bit tricky, but it can be obtained with the methods of ref.~\cite{Poisson:2009pwt}. The result can be expressed in terms of the $H_{i}$ functions (see \cite{Cano:2019ore}), and when evaluated on the stringy solution we get the following value for Hawking's temperature,

\begin{equation}\label{eq:temperature}
\begin{aligned}
T&=\frac{\sqrt{1-\chi^2}}{4\pi M\left(1+\sqrt{1-\chi^2}\right)}+\frac{\alpha'^2}{128\pi M^5}\bigg(\frac{73}{480}+\frac{2965 \chi ^2}{21504}\\
&+\frac{228349 \chi ^4}{1290240}+\frac{30077567 \chi ^6}{141926400}+{\cal O}(\chi^{8})\bigg)+\mathcal{O}(\alpha'^3)\, .
\end{aligned}
\end{equation}
In this case, we note that the corrections to the temperature are positive and they grow as we increase $\chi$. This is even more evident in fig.~\ref{fig:temperature}, where we show the temperature up to $\chi\sim0.9$ by using the $\mathcal{O}(\chi^{40})$ solution. 

At this point, we may wonder about the extremal limit $T=0$, but one has to be cautious with the $\alpha'$ and $\chi$ expansions, which are both singular for $\chi\sim 1$. To understand this, let us consider the corrections to the temperature of a near-extremal black hole. Assuming the extremal limit $T\rightarrow 0$ is regular, one can see that the temperature must be of the form
\begin{equation}\label{eq:Tne}
T^{\rm n.e.}\approx \frac{\sqrt{1+c\alpha'^2/M^4-\chi^2}}{4\pi M}\, ,\quad \text{when}\, \,\chi\sim 1\, ,
\end{equation}
where $c$ is certain dimensionless constant. Essentially, this formula expresses the fact that the temperature will tend to zero as a square root, $T\propto (\chi_{\rm ext}-\chi)^{1/2}$. This is the type of behavior one naturally expects near extremality even in the presence of higher-derivative corrections.\footnote{This has been explicitly observed, \textit{e.g.}, in the case of charged black holes with $\alpha'$ corrections \cite{Cano:2019ycn}.} If we now expand this expression in $\alpha'$ we get

\begin{equation}\label{eq:Tne2}
T^{\rm n.e.}\approx \frac{\sqrt{1-\chi^2}}{4\pi M}+\frac{c \alpha'^2}{8\pi M^5\sqrt{1-\chi^2}}+\mathcal{O}(\alpha'^3)\, .
\end{equation}
Thus, when expressed in this way, the correction to the temperature is divergent for $\chi=1$, but this divergence appears because the series expansion in $\alpha'$ is not valid anymore if $1-\chi^2\sim \alpha'^2/M^4$, so it is just an artifact.
Now we can ask if our correction to the temperature in eq.~\req{eq:temperature} behaves as $\sim 1/\sqrt{1-\chi^2}$ when $\chi\rightarrow 1$, but there are two caveats to this. In the first place, we do not know if the extremal limit is actually regular, and in fact studies of near-horizon geometries in EdGB and dCS theories suggest that it may not be \cite{Chen:2018jed}.\footnote{In particular, the near-horizon extremal geometries seem to be singular in EdGB gravity but they are regular for dCS theory \cite{Chen:2018jed}. However, the singularities in the EdGB case only appear at the poles of the horizon, and for instance the entropy is well-defined, so the thermodynamic properties probably behave regularly. It would be interesting to investigate what happens for the complete theory \req{eq:stringyfinal}.} In that case, it is not guaranteed one can then use \req{eq:Tne} for the near-extremal temperature. On the other hand, as we have already remarked, our series expansions do not converge for $\chi\sim 1$. Thus, the best we can do is to examine the correction to the temperature up to a large enough value of $\chi$ (in our case we can reach $\chi\sim 0.9$) and try to extrapolate the result for $\chi\sim 1$.
Let us define the dimensionless correction to the temperature by $T=T^{\rm Kerr}+\frac{\alpha'^2}{M^5}\Delta T$, so that $\Delta T$ is only a function of $\chi$. Then, our idea is to fit this function to an expression of the form

\begin{figure}[t!]
	\begin{center}
		\includegraphics[width=0.49\textwidth]{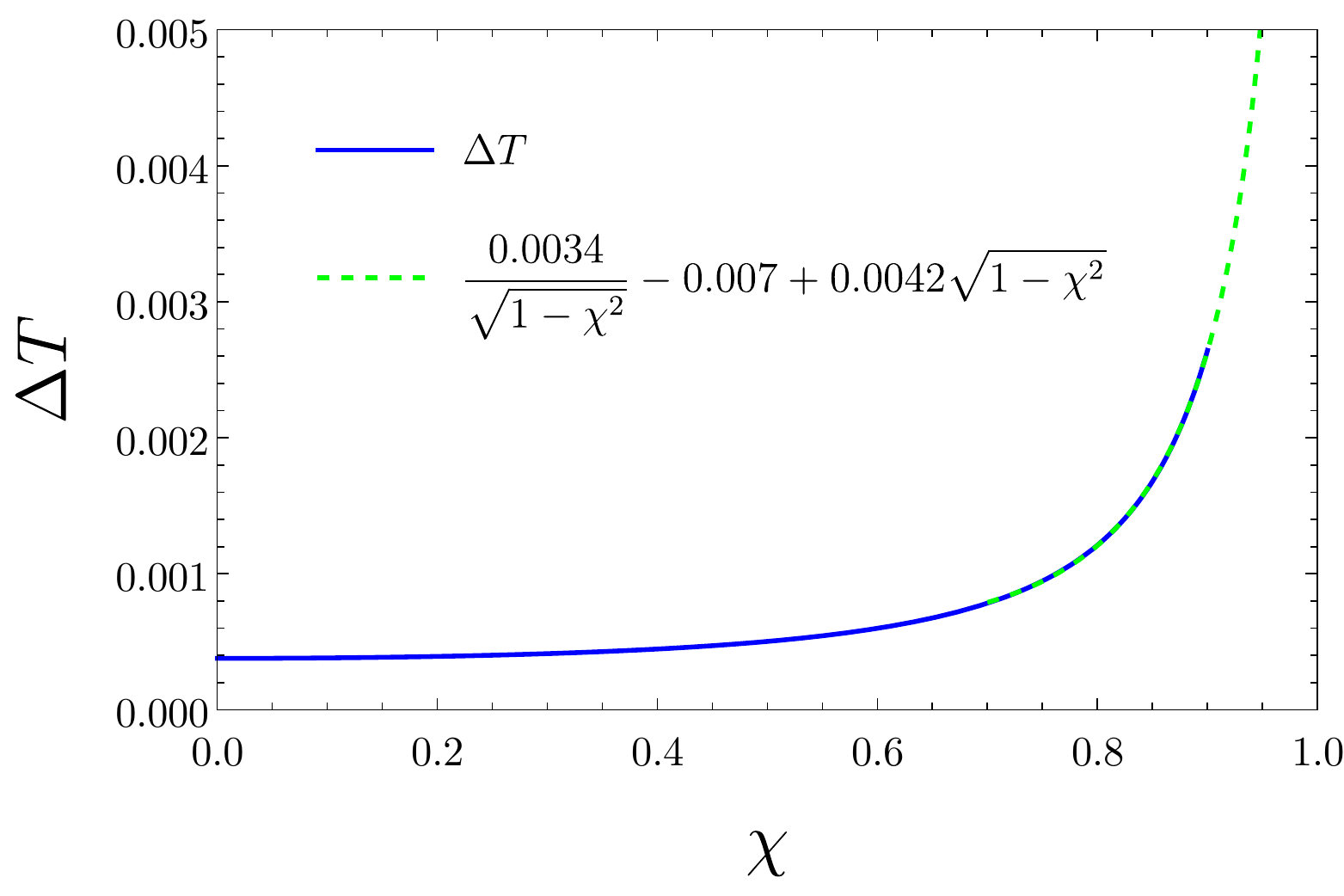}   
		\caption{Correction to the temperature $\Delta T=\frac{M^5}{\alpha'^2}(T-T^{\rm Kerr})$ as a function of the spin. We show the value of $\Delta T$ obtained with a series expansion to order $\chi^{40}$, which is accurate to $\chi\sim 0.9$, and the best fit of the form \req{eq:Tfit} performed in the interval $\chi\in [0.7,0.9]$. The results suggests that \req{eq:Tfit} indeed captures the behavior near $\chi\sim 1$.}
		\label{fig:deltaT}
	\end{center}
\end{figure}

\begin{equation}\label{eq:Tfit}
\Delta T=\frac{c_0}{\sqrt{1-\chi^2}}+c_1+c_2 \sqrt{1-\chi^2}+\ldots\, ,
\end{equation}
for values of $\chi$ as close to $1$ as we can. The result of this fit in the interval $\chi\in [0.7,0.9]$ --- in which our series expansion in \req{eq:temperature} is accurate --- is shown in fig.~\ref{fig:deltaT}. As we can observe, the fit with only three terms works very well, and it therefore suggests that the correction to the temperature really behaves as in \req{eq:Tne2}. We also obtain a value for the constant $c=8\pi c_0$, namely $c\approx 8\pi\times 0.0034\approx 0.086$, although it is hard to say how accurate this result is --- we would need to get closer to extremality to obtain a better estimation. On account of our previous discussion and equation \req{eq:Tne}, this would imply that the extremal limit is reached for
\begin{equation}
\chi_{\rm ext}=1+\frac{c\alpha'^2}{2M^4}+\mathcal{O}(\alpha'^3)>1\, .
\end{equation}
Determining the precise value of the spin at the extremal limit would require other methods than the ones we have applied here, but all of the evidence  points toward the existence of solutions with $\chi>1$.\footnote{As a matter of fact, this phenomenon has been observed for EdGB gravity \cite{Kleihaus:2011tg,Kleihaus:2015aje}, so it is not unexpected that the theory \req{eq:stringyfinal} shares this feature.}

Let us now turn our attention to the entropy of these black holes, which can be obtained by means of Wald formula \cite{Wald:1993nt, Iyer:1994ys}. This formula tells us that the entropy in a general diffeomorphism-invariant theory of gravity is given by 

\begin{equation}
{\cal S}=-2\pi \int_{\Sigma} d^2{x} \sqrt{|h|} \,\mathcal{E}^{\mu\nu\rho\sigma}_{R} \epsilon_{\mu\nu}\epsilon_{\rho\sigma}\, ,
\end{equation}
where $h$ and $\epsilon_{\mu\nu}$ are, respectively, the induced metric and the binormal (normalized so that $\epsilon_{\mu\nu}\epsilon^{\mu\nu}=-2$) of \emph{any}  cross-section of the horizon $\Sigma$, \footnote{In the original formula by Iyer and Wald, $\Sigma$ is assumed to be the bifurcation surface. However, it was later shown in \cite{Jacobson:1993vj} that the result is independent of which cross-section is chosen.} and 

\begin{equation}
\mathcal{E}_{R}^{\mu\nu\rho\sigma}=\frac{1}{\sqrt{|g|}}\frac{\delta S}{\delta R_{\mu\nu\rho\sigma}}\, ,
\end{equation}
is the variation of the four-dimensional action $S$ with respect to the Riemann tensor. Let us remark that this formula can only be applied to theories in which all the fields are tensors, \textit{i.e.}, there should be no fields with internal gauge freedom. This is an issue for the original form of the heterotic string effective action \req{heterotic1}, as the $B$-field contains Chern-Simons terms that are not (manifestly) invariant under diffeomorphisms. Historically, this has been a source of problems when dealing with the entropy of black holes in the heterotic theory (see for instance \cite{Sen:2005wa, Sahoo:2006pm, Tachikawa:2006sz, Faedo:2019xii, Edelstein:2019wzg, Cano:2021dyy} and references therein), that only recently has been rigorously understood \cite{Elgood:2020xwu,Elgood:2020nls}. However, in the form in which we expressed \req{eq:stringyfinal}, the Chern-Simons terms appear in a manifestly diff.-invariant form, and hence Wald's formula can be applied right away.

The variation of the action with respect to the curvature reads

\begin{equation}
\begin{aligned}
\mathcal{E}_{R}^{\mu\nu\rho\sigma}=&\frac{1}{16\pi}\left[g^{\mu [\rho}g^{\sigma] \nu}-\frac{\alpha'}{4}e^{-\phi}\tilde{\tilde{R}}^{\mu\nu\rho\sigma}\right.\\
&\left.-\frac{\alpha'}{8}\varphi ({\tilde R}^{\mu\nu\rho\sigma}+{\tilde R}^{\rho\sigma\mu\nu})\right]\, ,
\end{aligned}
\end{equation}
where 
\begin{equation}
\begin{aligned}
\tensor{\tilde{\tilde{R}}}{^{\mu\nu}_{\rho\sigma}}&=\frac{1}{4}\epsilon^{\mu\nu\alpha\beta}\epsilon_{\rho\sigma\lambda\tau}\tensor{R}{_{\alpha\beta}^{\lambda\tau}}\\
&=-\tensor{R}{^{\mu\nu}_{\rho\sigma}}+4\tensor{R}{^{[\mu}_{ [\rho}}\tensor{\delta}{^{\nu]}_{ \sigma]}}-R\tensor{\delta}{^{[\mu}_{ [\rho}}\tensor{\delta}{^{\nu]}_{ \sigma]}} 
\end{aligned}
\end{equation}
is the double dual of the Riemann tensor. However, it is known that the Gauss-Bonnet contribution to Wald's entropy can be simplified for stationary black holes, as the one at hand.  In fact, by decomposing the Riemann tensor at the horizon using the Gauss-Codazzi equations it is possible to show that 
\begin{equation}
\tilde{\tilde{R}}^{\mu\nu\rho\sigma}\epsilon_{\mu\nu}\epsilon_{\rho\sigma}=2 \mathcal{R}+\mathcal{O}(K^2)\, ,
\end{equation}
where $\mathcal{R}$ is the Ricci scalar of the induced metric on the horizon and $\mathcal{O}(K^2)$ are terms quadratic in the extrinsic curvatures of the horizon, that vanish for stationary black holes. Hence, the Wald formula reduces to the Jacobson-Myers result \cite{Jacobson:1993xs}, which takes a simpler form as it only involves intrinsic quantities. On the other hand, a similar analysis does not seem to lead to anything particularly illuminating for the Chern-Simons contribution. 

In sum, taking into account that $\epsilon_{\mu\nu}\epsilon^{\mu\nu}=-2$, we obtain the following general formula for the entropy of stationary black holes in the theory \req{eq:stringyfinal}, 

\begin{equation}\label{eq:Waldentropy}
{\cal S}=\frac{A_\Sigma}{4}+\frac{\alpha'}{32}\int_{\Sigma} d^2{x} \sqrt{|h|} \,\left[2e^{-\phi}  \mathcal{R}+\varphi {\tilde R}^{\mu\nu\rho\sigma}\epsilon_{\mu\nu}\epsilon_{\rho\sigma}\right] \, ,
\end{equation}
where $A_{\Sigma}$ denotes the area of any cross-section of the horizon.

Let us compute the different contributions in this formula, for instance, for a $t=$ constant slice of the horizon of our rotating black hole. The metric induced in this two-dimensional surface is the following:

\begin{equation}
\begin{aligned}
ds^2_{\Sigma}&=\left(1+H_{3}\right)\Sigma d\theta^2\\
&+\left(1+H_{4}\right)\frac{4M^2\rho^2_{+}\sin^2\theta}{\Sigma}d\phi^2\Big|_{\rho=\rho_{+}}\, .
\end{aligned}
\end{equation}
Therefore, the area is given by

\begin{equation}
\begin{aligned}
A_{\Sigma}=&4\pi M \rho_{+} \int_{-1}^{1} dx \left(1+\frac{H_3+H_4}{2}+{\cal O}(\alpha'^4)\right)\Bigg|_{\rho=\rho_{+}}\, ,
\end{aligned}
\end{equation}
and evaluating it in our solution we get

\begin{equation}
\begin{aligned}
A_{\Sigma}&=8\pi M^2\left(1+\sqrt{1-\chi^2}\right)-\frac{\pi\alpha'^2}{64M^2}\bigg(\frac{98}{5}+\frac{3959 \chi ^2}{560}\\
&+\frac{262877 \chi ^4}{100800}+\frac{7842301 \chi ^6}{31046400}+{\mathcal O}\left(\chi^{8}\right)\bigg)+\mathcal{O}(\alpha'^3)\, ,
\end{aligned}
\end{equation}
Let us remark at this point that the corrections to the area are strictly negative for every value of $\chi$; these black holes are more compact than their GR counterparts. 

Now we have to evaluate the integral in \req{eq:Waldentropy}. We note there is a first-order correction related to the Gauss-Bonnet term whose origin is topological. In fact, to first order in $\alpha'$ we have $e^{-\phi}=1-\phi$ and hence we have the term
\begin{equation}
\frac{\alpha'}{16}\int_{\Sigma} d^2{x} \sqrt{|h|} \mathcal{R}=\frac{\pi \alpha'}{4}\chi(\Sigma)=\frac{\pi \alpha'}{2}\, ,
\end{equation}
where in the first equality we used the Gauss-Bonnet theorem and in the second one we took into account that the Euler characteristic of a (topologically) spherical horizon is $\chi(\Sigma)=2$. On the other hand, there is no analogous topological contribution from the Chern-Simons term, because its integral vanishes.

Finally, we have to compute the dynamical contribution to the entropy by integrating the combination  $-2\phi \mathcal{R}+\varphi {\tilde R}^{\mu\nu\rho\sigma}\epsilon_{\mu\nu}\epsilon_{\rho\sigma}$. For that, we take into account that the scalar fields are already of order $\alpha'$ and therefore we can evaluate the curvatures on the Kerr metric. In addition, the appropriately normalized binormal to the horizon reads

\begin{equation}
\epsilon_{t\rho}=\frac{\rho_{+}^2+a^2 \cos^2\theta}{2M \rho_{+}}+\mathcal{O}(\alpha'^2)\,  .
\end{equation}

The evaluation of the integral yields the following result,


\begin{widetext}
\begin{equation}
\begin{aligned}
\frac{\alpha'}{8}\int_{\Sigma} d^2{x} \sqrt{|h|} \,\left[-2\phi\mathcal{R}+\varphi {\tilde R}^{\mu\nu\rho\sigma}\epsilon_{\mu\nu}\epsilon_{\rho\sigma}\right] =\frac{\pi \alpha'^2}{64M^2}\bigg[\frac{88}{3}+\frac{349 \chi ^2}{30}+\frac{27751 \chi ^4}{4200}+\frac{580801 \chi ^6}{141120}+{\cal O}(\chi^{8})\bigg]+\mathcal{O}(\alpha'^3)\, .
\end{aligned}
\end{equation}

\noindent
Then, putting all of the pieces together, we find the following result for the entropy at order $\alpha'^2$:

\begin{equation}\label{eq:entropy}
\begin{aligned}
{\cal S}=2\pi M^2\left(1+\sqrt{1-\chi^2}\right)+\frac{\pi \alpha' }{2}+\frac{\pi \alpha'^2}{64M^2}\bigg(\frac{73}{30}+\frac{7667 \chi ^2}{6720}+\frac{403147 \chi ^4}{403200}+\frac{271959 \chi ^6}{281600}+{\cal O}(\chi^{8})\bigg)+\mathcal{O}(\alpha'^3)\, .
\end{aligned}
\end{equation}
\end{widetext}
Interestingly, we observe that all of the corrections to the entropy are positive, despite the area term being corrected negatively. 

\begin{figure}[t!]
	\begin{center}
		\includegraphics[width=0.49\textwidth]{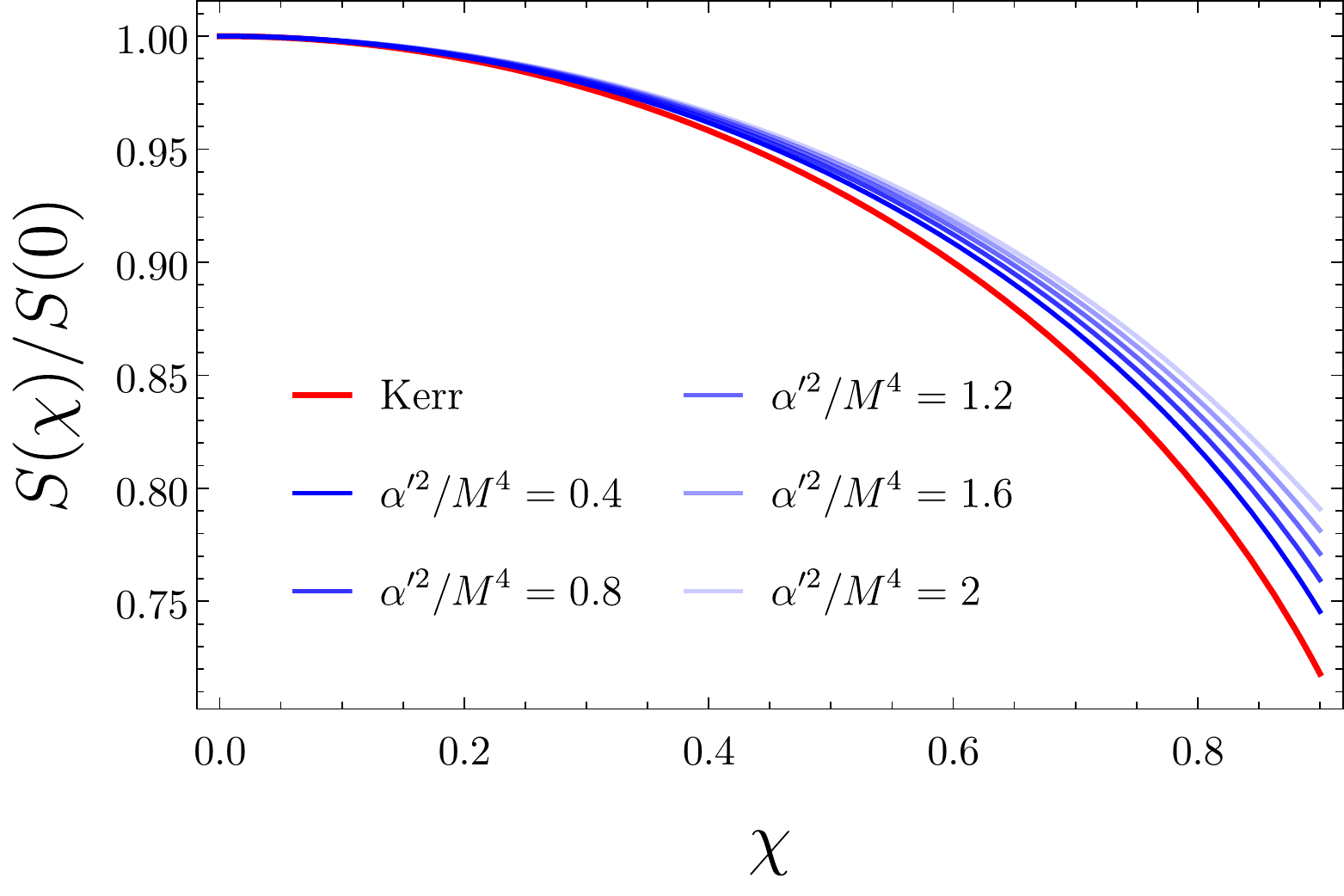}  
		\includegraphics[width=0.49\textwidth]{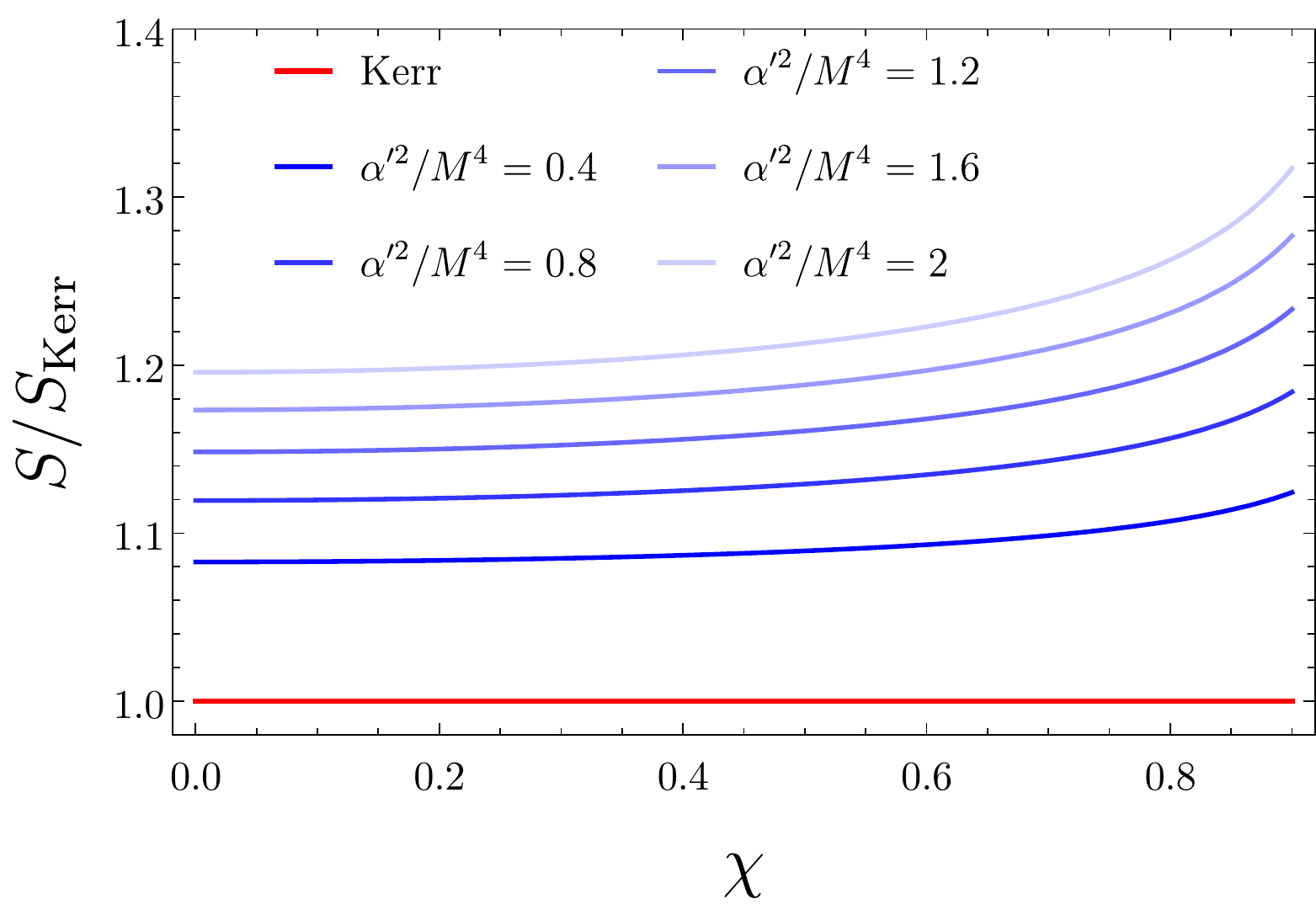}  
		\caption{Entropy of rotating stringy black holes as a function of the spin $\chi$. Top: entropy relative to the one of static black holes. Bottom: entropy relative to the one of a Kerr black hole. In red we show the Kerr result while the blue lines of different opacities correspond to different values of $\alpha'^2/M^4$. The entropy of the stringy black holes is larger than the one of Kerr, and the difference increases with $\chi$. }
		\label{fig:entropy}
	\end{center}
\end{figure}

Finally, we should check whether the first law of black-hole mechanics \cite{Bardeen:1973gs} is satisfied. For that, we must take into account that, in our solution, $M$ represents indeed the physical mass of the black hole, while the angular momentum is $J=\chi M^2$. Then, using \req{eq:omega}, \req{eq:temperature} and \req{eq:entropy}, we observe that the first law 

\begin{equation}
\delta M=T \delta {\cal S}+ \Omega\delta J\, ,
\end{equation}
holds at order $\alpha'^2$. Although here we are only showing the solution at order $\chi^6$, we have checked that the first law is satisfied for all the terms in the $\chi$-expansion that we are able to compute --- up to $\sim \chi^{40}$.  This is a remarkably strong test on the validity of our computations, but also an interesting check on the validity of the first law of black hole mechanics for heterotic string theory.
We offer two different visualizations of the entropy in fig.~\ref{fig:entropy}. In the top plot we show  the entropy of rotating black holes relative to that of static ones, and we observe that, while the entropy always decreases with the spin, it does so more slowly when the $\alpha'$ corrections are taken into account. In the second plot we graph the ratio between the entropy of the corrected black holes and the one of the Kerr black hole. We see that, not only the entropy of the stringy black holes is larger than the Kerr one, but the difference increases as we turn on the spin. All these effects seem to become more drastic as we approach $\chi\sim 1$, so it would be interesting to explore what happens in the near-extremal case, that is not accessible with our analysis. 

\section{Conclusions}\label{sec:conclusions}
We have performed a dimensional reduction and truncation of the heterotic string effective action down to four dimensions. The resulting theory can be expressed in a very appealing form as given by eq.~\req{eq:stringyfinal}, which is a simple generalization of the well studied EdGB and dCS models. We have observed that the appearance of the Gauss-Bonnet and Pontryagin densities and no other higher-derivative terms is in fact a natural and non-trivial prediction of heterotic string theory. Interestingly, the 4-dimensional action resembles that of a standard axidilaton model (see \req{eq:stringyaxidilaton}) where the curvature plays the form of the field strength of a gauge field.  It is therefore tantalizing to speculate whether such an action could contains a sort of $\mathrm{SL}(2,\mathbb{R})$ duality symmetry involving the curvature. If one could make sense of such a symmetry, this would be a very interesting way to constrain the additional $\alpha'$ corrections in this minimal setup that only contains the metric and the axidilaton. 

Expressed as in eq.~\req{eq:stringyfinal}, the action contains $\mathcal{O}(\alpha'^2)$ corrections, which we show explicitly in the Appendix \ref{app:app:alpha2}. However, these terms become of higher order if one is interested in corrections to vacuum GR solutions (with trivial scalars at zeroth order).  Therefore, the simple action \req{eq:stringyfinal} can be used to compute consistently all of the corrections to Ricci flat solutions to order $\alpha'^2$. 
Then, using the methods of \cite{Cano:2019ore} we have obtained the perturbative $\mathcal{\alpha'}^2$ corrections to the Kerr background expressed as a power series in the dimensionless spin parameter $\chi=J/M^2$. This approach allows one to study non-extremal black holes, but with enough terms in the series one can get close to extremality $\chi\sim 1$. In our case, we managed to obtain the solution to order $\chi^{40}$, which is accurate up to $\chi\sim 0.9$. 
We have focused on the thermodynamic properties of these stringy rotating black holes, and in particular we have obtained their entropy at order $\alpha'^2$. We have found that, for a fixed angular momentum and mass, these black holes spin more slowly, are more compact, are hotter and are more entropic than their Kerr counterparts. In addition all these effects become more relevant as the spin increases.  We have also checked that the first law of black hole mechanics holds at order $\alpha'^2$, which is a very strong test of our calculations. 

The positivity of the corrections to the entropy is line with general expectations, as it indicates that the underlying quantum theory contains more degrees of freedom that only become active at higher energies. On the other hand, the fact that the temperature is higher than for Kerr black holes suggests that the extremal value of the angular momentum will be corrected positively. We cannot obtain a precise value for the extremality bound with our current approach, but as discussed in section~\ref{sec:kerrbh}, everything points towards the fact that extremality is modified as
\begin{equation}
\frac{J}{M^2}\Big|_{\rm ext}=1+c\frac{\alpha'^2}{M^4}+\mathcal{O}(\alpha'^2)\, ,
\end{equation}
for a positive constant $c$. Interestingly, this is reminiscent of a form of the weak gravity conjecture according to which the corrections to the charge-to-mass ratio of extremal charged black holes should be positive in string theory \cite{Kats:2006xp,Hamada:2018dde,Cheung:2018cwt}. The positivity of the corrections to the entropy is also connected with this statement \cite{Goon:2019faz}. One of the reasons behind this conjecture is the requirement that extremal black holes should be able to discharge, which is possible if $Q/M$ is modified positively by the higher-derivative corrections.
In the case of rotating black holes, there does not seem to be an obvious reason why $J/M^2$ should be corrected positively, as the angular momentum can always be radiated away.\footnote{See nevertheless refs.~\cite{Cheung:2019cwi,Aalsma:2020duv} on possible extensions of the weak gravity conjecture for spinning black holes.} Our results suggest that  $J/M^2>1$ anyway, in complete analogy with the charged case.

It would certainly be interesting to better understand the properties of extremal and near-extremal black holes in the theory \req{eq:stringyfinal}. One possibility to to simplify the problem would be to study the near-horizon geometry of extremal black holes, which is itself a solution with enhanced symmetry to the equations of motion \cite{Chen:2018jed,Cano:2019ozf}. The near-horizon geometry allows one to obtain the entropy of extremal black holes, but the expression obtained is meaningless unless it can be written in terms of physical quantities.\footnote{In ref.~\cite{Chen:2018jed}, the entropy of extremal black holes in EdGB and dCS theories is computed. The result is expressed in terms of a parameter $M$ that is claimed to be the mass, but the higher-derivative corrections could renormalize the mass, thus changing the interpretation of $M$. This cannot be seen in the near-horizon geometry alone.} One could compute the angular momentum by making use of generalized Komar integrals \cite{Ortin:2021ade}, in whose case one would obtain a physically meaningful relation ${\cal S}(J)$. However, one cannot identify the mass of the black hole from the near-horizon geometry.  
Thus, in order to compute the extremality bound one would need to study the global geometry of extremal black holes, which probably can only be accessed numerically --- see \cite{McNees:2015srl} though.\footnote{Ref.~\cite{Reall:2019sah} provides another way to obtain analytically the corrections to the thermodynamic properties of Kerr black holes. However, that method is only applicable to first-order corrections, while in our case the thermodynamic quantities are modified at second order. }

Finally, it would be interesting to study the solutions to \req{eq:stringyfinal} even in a non-perturbative fashion. Although strictly speaking one would lose contact with string theory (because the stringy action contains more terms besides those in \req{eq:stringyfinal}), there are at least two reasons to do this. 
On the one hand, the theory \req{eq:stringyfinal} is, on its own, a well-motivated and interesting model. As it is known, the Gauss-Bonnet invariant leads to second order equations of motion, and the Pontryagin density has also reduced-order equations, namely, of third order (instead of fourth order, which is the case for typical higher-curvature terms). Thus, the theory might have a chance to lead to a well-posed dynamical evolution, which would be worth exploring \cite{Delsate:2014hba}. 
On the other hand, non-perturbative solutions may offer new phenomena that are not seen when performing a perturbative expansion in $\alpha'$, hence their interest. Also, in that case the coupling between the axion and the dilaton and the presence of both Gauss-Bonnet and Pontryagin densities will make the non-perturbative solutions substantially different to their EdGB  \cite{Kleihaus:2011tg,Kleihaus:2015aje} and dCS \cite{Delsate:2018ome} counterparts, leading perhaps to interesting new properties. 

\vspace{0.4cm}
\begin{acknowledgments}   
\textbf{\textit{Acknowledgments.}}
We would like to thank Tom\'as Ort\'in, Carlos Shahbazi,  Nikolay Bobev, Nicol\'as Yunes, Augusto Sagnotti, Guiseppe Dibitetto, Davide Cassani, Lars Aalsma and Brian McPeak for comments and discussions. The work of PAC is supported by a postdoctoral fellowship from the Research Foundation - Flanders (FWO grant 12ZH121N). AR is supported by a postdoctoral fellowship associated to the MIUR-PRIN contract 2017CC72MK003.  
\end{acknowledgments}

\appendix
\onecolumngrid \vspace{1.5cm}

\appendix
\section{Field redefinitions and the $\mathcal{O}(\alpha'^2)$ terms}\label{app:app:alpha2}

As we have discussed in the main text, the dualization of the Kalb-Ramond 2-form $B$ introduces ${\cal O}(\alpha'^2)$ terms that we have  however ignored in \eqref{heterotic3}. The effective Lagrangian including these terms is given by 

\begin{align}
\mathcal{L}=e^{-2(\hat\phi-\hat\phi_{\infty})}\Big(\bar R-4(\partial \hat\phi)^2\Big)+\frac{1}{2}e^{2(\hat\phi-\hat\phi_{\infty})}(\partial\varphi)^2+\frac{\alpha'}{8}\mathcal{L}_{R^2}\Big|_{H^{(0)}}-\frac{\alpha'^2}{12}e^{-2(\hat \phi-\hat \phi_{\infty})}H^{(1)}_{\mu\nu\rho}H^{(1)}{}^{\mu\nu\rho}+\mathcal{O}(\alpha'^3)\, ,
\label{heteroticalpha2terms}
\end{align}
where 

\begin{align}\label{H^{(1)}}
H^{(1)\mu\nu\rho}=&-e^{2(\hat \phi-\hat \phi_{\infty})}\frac{3}{4}\frac{\delta {\cal L}_{R^2}}{\delta H_{\mu\nu\rho}}\Bigg|_{H^{(0)}}=-e^{2(\hat \phi-\hat \phi_{\infty})}\frac{3}{2}\left\{{\bar\nabla}_{\sigma}\left(e^{-2(\hat \phi-\hat \phi_{\infty})}{\bar  R}^{(0)}_{(-)}{}^{\sigma [\mu\nu\rho]}-\varphi {\tilde {\bar R}}^{(0)}_{(-)}{}^{\sigma [\mu\nu\rho]}\right)\right.\\
+&\left. \left(e^{-2(\hat \phi-\hat \phi_{\infty})}{\bar  R}^{(0)}_{(-)}{}^{\alpha [\mu\rho|\beta}-\varphi {\tilde {\bar R}}^{(0)}_{(-)}{}^{\alpha [\mu\rho|\beta}\right) H^{(0)}_{\alpha\beta}{}^{\nu]}\right\}\,, \nonumber
\end{align}
and

\begin{equation}
{\bar  R}^{(0)}_{(-)}{}_{\mu\nu\rho\sigma}\equiv {\bar  R}_{(-)}{}_{\mu\nu\rho\sigma}\Big|_{H^{(0)}}\, .
\end{equation}
One could now use \eqref{H0} and \eqref{Rminus} in order to find the expression of $H^{(1)}$ in terms of the Riemann tensor and the scalars, and then plug it back into \eqref{heteroticalpha2terms} to write down the $\alpha'^2$ corrections using these variables. This is however a long calculation that we are going to avoid since we do not need to know explicitly these terms. All we have to check, before ignoring them once for all, is that they will not induce $\alpha'^2$ corrections to vacuum solutions of GR. This is not difficult to see, as one can check by using the Bianchi identities of the Riemann tensor (namely, ${\bar R}_{\mu[\nu\rho\sigma]}=0$ and ${\bar\nabla}^{\sigma}\tilde {\bar R}_{\sigma \mu\nu\rho}=0$) in the first term of \eqref{H^{(1)}} that all the terms entering in the expression for $H^{(1)}$ contain derivatives of the scalars. This implies that the $\alpha'^2$ terms in the action actually become effectively of order ${\cal O}(\alpha'^4)$ when this action is used to compute corrections to vacuum solutions of GR. Hence, one can simply ignore them if we are just interested in the leading ${\cal O}(\alpha'^2)$ corrections.

Let us now give further details of the field redefinitions that one has to perform in order to cancel all the terms contained in ${\cal L}'$. Let us recall that, in terms of $\phi=2(\hat\phi-\hat\phi_{\infty})$, the Lagrangian expressed in the modified Einstein frame reads

\begin{align}\label{LagEinst}
\mathcal{L}=& R+\frac{1}{2}(\partial\phi)^2+\frac{1}{2}e^{2\phi}(\partial \varphi)^2+\frac{\alpha'}{8}\left(e^{-\phi}\mathcal{X}_{4}-\varphi R_{\mu\nu\rho\sigma}\tilde R^{\mu\nu\rho\sigma}+\mathcal{L}'\right)+\mathcal{O}(\alpha'^2)\, ,
\end{align}
where

\begin{equation}
\begin{aligned}
\mathcal{L}'=e^{-\phi}\Big[4\mathcal{E}_{\mu\nu}\mathcal{E}^{\mu\nu}-\mathcal{E}^2+2\E\E_{\phi}
+3\mathcal{E}_{\phi}^2-3\E_{\varphi}^2+\E_{\phi}(A^2-(\partial\phi)^2)-2\E_{\varphi}\partial_{\mu}\phi A^{\mu}\Big]\, ,
\end{aligned}
\end{equation}
and where 

\begin{align}
\mathcal{E}_{\mu\nu}&=R_{\mu\nu}+\frac{1}{2}\partial_{\mu}\phi\partial_{\nu}\phi+\frac{1}{2}A_{\mu}A_{\nu}\, ,\quad \E=\E_{\mu\nu}g^{\mu\nu}\\
\mathcal{E}_{\phi}&=\nabla^2\phi-A^2\, ,\\
\mathcal{E}_{\varphi}&=\nabla_{\mu}A^{\mu}+\partial_{\mu}\phi A^{\mu}\, ,
\end{align}
and we recall that $A_{\mu}=e^{\phi}\partial_{\mu}\varphi$. We then perform field redefinitions of order $\alpha'$, 

\begin{equation}
g_{\mu\nu}\rightarrow g_{\mu\nu} +\frac{\alpha'}{8} \Delta_{\mu\nu}\, ,\quad \phi\rightarrow \phi+\frac{\alpha'}{8} \Delta {\phi}\, ,\quad \varphi\rightarrow \varphi+\frac{\alpha'}{8} \Delta {\varphi}\ .
\end{equation}
It is not difficult to see that, up to boundary terms that we discard, these redefinitions have the following effect in the Lagrangian,

\begin{equation}
\mathcal{L}\rightarrow \mathcal{L}-\frac{\alpha'}{8}\left[\left(\E^{\mu\nu}-\frac{1}{2}\E g^{\mu\nu}\right)\Delta_{\mu\nu}+\E_{\phi}\Delta_{\phi}+\E_{\varphi}\Delta_{\varphi}\right]+\mathcal{O}(\alpha'^2)\, .
\end{equation}
Therefore, we can use these redefinitions to cancel all of the terms in $\mathcal{L}'$. This is for instance achieved by
\begin{align}
\Delta_{\mu\nu}&=4\E_{\mu\nu}-(\E+2\E_{\phi})g_{\mu\nu}\, ,\\
\Delta_{\phi}&=3\E_{\phi}+A^2-(\partial\phi)^2\, ,\\
\Delta_{\varphi}&=-3\E_{\varphi}-2\partial_{\mu}\phi A^{\mu}\, ,
\end{align}
the choice being not unique. Notice that the redefinition not only introduces $\mathcal{O}(\alpha')$ terms canceling $\mathcal{L}'$, but also introduces an infinite tower of $\alpha'^n$ terms, that includes in particular $\alpha'^2$ corrections, besides those already present originally in \req{LagEinst}. These new terms are proportional to the $\Delta$ shifts. In the case of $\Delta_{\mu\nu}$, this quantity is also proportional to the zeroth-order equations of motion, and therefore the  $\mathcal{O}(\alpha'^2)$ terms generated by the transformation $\Delta_{\mu\nu}$ can be removed by a new field redefinition of order $\alpha'^2$. The same reasoning applies to the part of $\Delta_{\phi}$ and $\Delta_{\varphi}$ that is proportional to the zeroth-order EOMs. Therefore, the only $\alpha'^2$ terms that we cannot \textit{a priori} get rid of are those related to the transformations
\begin{align}
\tilde\Delta_{\phi}=A^2-(\partial\phi)^2\, ,\quad
\tilde\Delta_{\varphi}=-2\partial_{\mu}\phi A^{\mu}\, .
\end{align}
Since these are related to redefinitions of the scalars, it is easy to compute their effect, and we see that they generate the following contribution to the six-derivative Lagrangian

\begin{equation}
\begin{aligned}
\mathcal{L}_{(6)}\supset\frac{\alpha'^2}{64}\left[\frac{1}{2}(\partial\tilde\Delta_{\phi})^2+\frac{1}{2}e^{2\phi}(\partial\tilde\Delta_{\varphi})^2+2e^{2\phi}\tilde\Delta_{\phi}\partial_{\mu}{\varphi}\partial^{\mu}\tilde\Delta_{\varphi}+e^{2\phi}\tilde\Delta_{\phi}^2(\partial\varphi)^2-e^{-\phi}\tilde\Delta_{\phi}\mathcal{X}_{4}-\tilde\Delta_{\varphi}R_{\mu\nu\rho\sigma}\tilde R^{\mu\nu\rho\sigma}\right]\, .
\end{aligned}
\end{equation} 
We suspect that this Lagrangian, together with the $(H^{(1)})^2$ contribution in \req{heteroticalpha2terms}, can be simplified by integrating by parts and using additional $\mathcal{O}(\alpha'^2)$ field redefinitions. However, for the purposes of this paper we only need to note that the quantities $\tilde\Delta_{\phi,\varphi}$ are proportional to the square of $\partial\phi$ and $\partial\varphi$. Therefore, for corrections over vacuum GR solutions we have $\tilde\Delta_{\phi,\varphi}\sim \mathcal{O}(\alpha'^2)$ and in those cases the Lagrangian $\mathcal{L}_{(6)}$ actually contributes to the equations of motion at order $\mathcal{O}(\alpha'^4)$. Hence, the first-order Lagrangian \req{eq:stringyfinal} already provides all of the $\alpha'^2$ corrections to the vacuum GR solutions. 

\section{The corrected solution}\label{app:thesolution}

The scalar fields, to order $\chi^4$, are given by

\begin{eqnarray}
\phi&=&\alpha'\left\{-\frac{M}{3 \rho ^3}-\frac{1}{4 \rho ^2}-\frac{1}{4 M \rho }+\left[x^2\left(\frac{12 M^3}{5 \rho ^5}+\frac{21 M^2}{20 \rho ^4}+\frac{7 M}{20 \rho ^3}\right)+\frac{M^2}{40 \rho ^4}+\frac{M}{20 \rho ^3}+\frac{1}{16 \rho ^2}+\frac{1}{16 M \rho }\right]\chi^2\right. \nonumber\\
&+&\left[x^4\left(-\frac{45 M^5}{7 \rho ^7}-\frac{55 M^4}{28 \rho ^6}-\frac{11 M^3}{28 \rho ^5}\right)+x^2\left(-\frac{M^4}{14 \rho ^6}-\frac{3 M^3}{35 \rho ^5}-\frac{3 M^2}{56 \rho ^4}-\frac{M}{56 \rho ^3}\right)+\frac{M^3}{140 \rho ^5}+\frac{M^2}{56 \rho ^4}+\frac{3 M}{112 \rho ^3}\right.\nonumber\\
&+&\left.\left.\frac{1}{32 \rho ^2}+\frac{1}{32 M \rho }\right]\chi^4+{\cal O}(\chi^6)\right\}+{\cal O}(\alpha'^2)\,,\\
&&\nonumber\\
\varphi&=&\alpha'\left\{x \left(-\frac{9 M^2 }{8 \rho ^4}-\frac{5 M }{8 \rho ^3}-\frac{5 }{16 \rho ^2}\right)\chi +\left[x^3\left(\frac{25 M^4}{6 \rho ^6}+\frac{3 M^3}{2 \rho ^5}+\frac{3 M^2}{8 \rho ^4}\right)\right.\right.\nonumber\\
&+& \left. \left. x\left(\frac{M^3 }{20 \rho ^5}+\frac{3 M^2 }{40 \rho ^4}+\frac{M }{16 \rho ^3}+\frac{1}{32 \rho ^2}\right)\right]\chi ^3+{\cal O}(\chi^5) \right\}+{\cal O}(\alpha'^2)\, ,
\end{eqnarray}
where we recall that $x=\cos\theta$. The expressions for the $H_{i}$ functions are lengthier, and hence we only show here their expressions to order $\chi^2$,

\begin{eqnarray}\nonumber\\
H_1&=&\alpha'^2\left\{\frac{13 M^3}{22 \rho ^7}+\frac{7 M^2}{660 \rho ^6}+\frac{107 M}{18480 \rho ^5}-\frac{1601}{12320 \rho ^4}-\frac{61}{12320 M \rho ^3}-\frac{1117}{73920 M^2 \rho ^2}+\frac{1117}{73920 M^3 \rho }\right.\nonumber\\
&+&\left[x^2 \left(-\frac{15289 M^5}{5280 \rho ^9}+\frac{2867383 M^4}{2242240 \rho ^8}+\frac{153473 M^3}{240240 \rho ^7}+\frac{576768737 M^2}{1009008000 \rho ^6}+\frac{20236591 M}{2018016000 \rho ^5}+\frac{20614157}{576576000 \rho ^4}\right.\right.\nonumber\\
&-&\left.\left.\frac{2878313}{52416000 M \rho ^3}\right)-\frac{1744903 M^4}{1681680 \rho ^8}-\frac{2088607 M^3}{3363360 \rho ^7}-\frac{324773297 M^2}{1009008000 \rho ^6}+\frac{349537 M}{42042000 \rho ^5}+\frac{11853323}{288288000 \rho ^4}\right.\nonumber\\
&+&\left.\left.\frac{2844323}{288288000 M \rho ^3}-\frac{883349}{230630400 M^2 \rho ^2}+\frac{883349}{230630400 M^3 \rho }\right]\chi^2+{\cal O}(\chi^4)\right\}+{\cal O}(\alpha'^3)\, ,\\
&&\nonumber\\
H_2&=&\alpha'^2\left\{\frac{13 M^2}{44 \rho ^6}-\frac{4871 M}{21120 \rho ^5}-\frac{1061}{7392 \rho ^4}-\frac{10963}{73920 M \rho ^3}-\frac{337}{73920 M^2 \rho ^2}-\frac{1117}{147840 M^3 \rho }+\frac{1117}{147840 M^4}\right.\nonumber\\
&+&\left[x^2 \left(-\frac{12169 M^4}{10560 \rho ^8}+\frac{9398303 M^3}{6726720 \rho ^7}+\frac{6719353 M^2}{13453440 \rho ^6}+\frac{9627139 M}{44352000 \rho ^5}-\frac{799998211}{16144128000 \rho ^4}-\frac{119427883}{6457651200 M \rho ^3}\right.\right.\nonumber\\
&-&\left.\frac{35886187}{1076275200 M^2 \rho ^2}\right)-\frac{1744903 M^3}{3363360 \rho ^7}-\frac{11445271 M^2}{40360320 \rho ^6}-\frac{527219389 M}{4036032000 \rho ^5}+\frac{136836577}{5381376000 \rho ^4}+\frac{1014817483}{32288256000 M \rho ^3}\nonumber\\
&+&\left.\left.\frac{99396923}{16144128000 M^2 \rho ^2}-\frac{883349}{461260800 M^3 \rho }+\frac{883349}{461260800 M^4}\right]\chi^2+{\cal O}(\chi^4)\right\}+{\cal O}(\alpha'^3)\, ,\\
&&\nonumber\\
H_3&=&\alpha'^2\left\{-\frac{23 M^2}{132 \rho ^6}-\frac{73 M}{660 \rho ^5}-\frac{551}{7392 \rho ^4}-\frac{101}{18480 M \rho ^3}-\frac{19}{73920 M^2 \rho ^2}+\frac{1117}{73920 M^3 \rho }-\frac{1117}{73920 M^4}\right.\nonumber\\
&+&\left[x^2 \left(\frac{7153 M^4}{10560 \rho ^8}+\frac{2604257 M^3}{13453440 \rho ^7}+\frac{54877 M^2}{2446080 \rho ^6}-\frac{104823353 M}{1009008000 \rho ^5}-\frac{907967}{12812800 \rho ^4}-\frac{241183}{4576000 M \rho ^3}\right.\right.\nonumber\\
&+&\left.\frac{254477}{230630400 M^2 \rho ^2}\right)+\frac{525947 M^3}{13453440 \rho ^7}+\frac{1765081 M^2}{26906880 \rho ^6}+\frac{6740833 M}{91728000 \rho ^5}+\frac{14704687}{288288000 \rho ^4}+\frac{5882663}{288288000 M \rho ^3}\nonumber\\
&+&\left.\left.\frac{50639}{9609600 M^2 \rho ^2}+\frac{883349}{230630400 M^3 \rho }-\frac{883349}{230630400 M^4}\right]\chi^2+{\cal O}(\chi^4)\right\}+{\cal O}(\alpha'^3)\, ,\\
&&\nonumber\\
H_4&=&\alpha'^2\left\{-\frac{23 M^2}{132 \rho ^6}-\frac{73 M}{660 \rho ^5}-\frac{551}{7392 \rho ^4}-\frac{101}{18480 M \rho ^3}-\frac{19}{73920 M^2 \rho ^2}+\frac{1117}{73920 M^3 \rho }-\frac{1117}{73920 M^4}\right.\nonumber\\
&+&\left[x^2\left(-\frac{31 M^5}{33 \rho ^9}+\frac{4277 M^4}{3520 \rho ^8}+\frac{1173329 M^3}{2242240 \rho ^7}+\frac{5905531 M^2}{13453440 \rho ^6}-\frac{741859 M}{22932000 \rho ^5}+\frac{11743751}{576576000 \rho ^4}\right.\right.\nonumber\\
&-&\left.\frac{2878313}{52416000 M \rho ^3}\right)+\frac{31 M^5}{33 \rho ^9}-\frac{2839 M^4}{5280 \rho ^8}-\frac{390977 M^3}{1345344 \rho ^7}-\frac{429197 M^2}{1223040 \rho ^6}+\frac{983803 M}{504504000 \rho ^5}-\frac{644247}{16016000 \rho ^4}\nonumber\\
&+&\left.\left.\frac{13037711}{576576000 M \rho ^3}+\frac{1469813}{230630400 M^2 \rho ^2}+\frac{883349}{230630400 M^3 \rho }-\frac{883349}{230630400 M^4}\right]\chi^2+{\cal O}(\chi^4)\right\}+{\cal O}(\alpha'^3)\, .
\end{eqnarray}

\bibliographystyle{apsrev4-1} 
\bibliography{StringGravity}

\end{document}
%